\documentclass[pra,preprint,superscriptaddress,floatfix,showpacs, 11pt]{revtex4-1}
\usepackage{hyperref} % For hyperlinks
\usepackage{amsmath,amsfonts,amssymb}
\usepackage{graphicx}
\usepackage{xcolor}
\usepackage{makecell}
\usepackage{natbib}

\begin{document}
\title{Inferring rotations using a bosonic Josephson junction}
\author{Rhombik Roy}
\email{rroy@campus.haifa.ac.il}
\affiliation{Department of Physics, University of Haifa, Haifa 3498838, Israel}
\affiliation{Haifa Research Center for Theoretical Physics and Astrophysics, University of Haifa,
Haifa 3498838, Israel}
\author{Ofir E. Alon}
\affiliation{Department of Physics, University of Haifa, Haifa 3498838, Israel}
\affiliation{Haifa Research Center for Theoretical Physics and Astrophysics, University of Haifa,
Haifa 3498838, Israel}

%\date{\today}

\begin{abstract} 
Rotation and quantum tunneling are fundamental concepts in physics, and their interplay in the ultracold atomic systems is of particular interest. In this theoretical work, we explore how tunneling dynamics in a bosonic Josephson junction are modified when the system is placed in a rotating, non-inertial frame. We show that the tunneling dynamics of ultracold bosons in a two-dimensional double-well potential offer an alternative pathway for inferring the rotation frequency. Using the mean-field and many-body analyses, we demonstrate that rotation strongly modifies the tunneling time period as well as the momentum and angular momentum dynamics. 
When the rotation axis passes through the center of the double well, the observables show distinct dynamical responses with increasing rotation frequency, enabling the rotation frequency to be assessed from changes in the tunneling dynamics. Specifically, the tunneling period and the time-averaged angular momentum increase exponentially, while the amplitude of the transverse momentum increases linearly.
When the potential is displaced from the rotation axis, the rotation induces asymmetric tunneling and partial self-trapping, allowing both the rotation frequency and the displacement to be inferred. In this scenario, self-trapping increases exponentially with both the displacement and the rotation frequency, while the time-averaged angular momentum increases linearly with the rotation frequency and quadratically with the displacement. 
We further show that for an off-centered double well, the tunneling dynamics exhibit a pronounced orientation dependence, enabling the orientation of the double well to be inferred from the observed dynamics. In this case, both the self-trapping and the time-averaged angular momentum display a Gaussian dependence on the orientation. 
The many-body analysis further shows that the depletion dynamics are strongly influenced by rotation, providing an additional tool for assessing the rotation frequency. In particular, the depletion exhibits an exponential dependence on both the rotation frequency and the orientation of the potential. 
Finally, we study the effect of time-dependent rotation in which the double well is gradually set into motion in the laboratory frame and identify distinct dynamical signatures that depend sensitively on the switching time. Together, these results establish a comprehensive framework for inferring the rotation frequency, radial displacement, and orientation directly from the tunneling dynamics.
\end{abstract}

\maketitle

\section{Introduction}\label{intro} 
The introduction of rotation into ultracold atomic systems alters their quantum behavior, giving rise to rich and distinctive dynamical responses. It enables the creation of quantized vortices, vortex lattices, and a variety of topological states~\cite{ref1,fetter}, and it provides a versatile platform to simulate phenomena ranging from superfluidity and quantum turbulence to artificial gauge fields and fractional quantum Hall physics~\cite{ref2,ref3,ref4,ref5,ref6,gammal_rotation}. By controlling rotation, experiments can probe angular-momentum quantization, out-of-equilibrium dynamics, and collective excitations in Bose–Einstein condensates and Fermi gases, making rotation a key tool for probing quantum behavior and for engineering novel many-body states~\cite{phantom_angular_axel,ref7,ref8,axel_pelster1,axel_pelster2,vittorio}.
Beyond this, the high degree of control achieved in rotating ultracold systems also opens promising avenues for applications in quantum technologies, including high-precision rotation sensing and inertial navigation, as well as matter-wave interferometry for probing fundamental physics~\cite{main_ref1,main_ref2,main_ref3}. 

Rotation is unique among physical quantities in that it can be measured intrinsically, independent of any external inertial frame~\cite{app1}.  Rotation sensing is essential for precisely tracking angular motion, with applications ranging from inertial navigation and vehicle guidance to spacecraft  stabilization~\cite{app2}.
Accurate rotation sensing also plays an important role in geophysics, structural monitoring, consumer electronics, and advanced quantum technologies~\cite{app3,app4}. Consequently, the development of compact and accurate rotation sensors is thus crucial for both practical applications and precision quantum experiments.

For many years, ring-laser and fiber-optic gyroscopes based on Sagnac effect have set the standard for precision rotation sensing, where sensitivity improves with the size of the enclosed optical path~\cite{app11,app12,app14,app15,app16}. Although these devices have become smaller over time, they still face fundamental limits due to their physical size~\cite{app21,app22}. At the other end, compact MEMS (micro-electro-mechanical system) gyroscopes have become ubiquitous in consumer electronics, but their accuracy remains far below that of optical systems~\cite{app31,app32,app33,app34}. In recent years, different paths have begun to reshape the landscape of precision rotation sensing~\cite{app41,app42,app43,app44}. Among them, quantum interferometry has evolved into a versatile framework that now includes electrons, neutrons, cold atoms, superfluids, and Bose–Einstein condensates~\cite{app51,app52,app53,app54,app55,app56,app57}. This development has opened new possibilities for compact rotation sensors with quantum-enhanced sensitivity.  

In this context, bosonic Josephson junctions (BJJs) may offer a viable route toward rotation sensing. BJJs provide an ideal platform for exploring macroscopic quantum phenomena~\cite{Ulbricht,intro1,intro3}. Extensive studies have examined BJJ dynamics in both bosonic~\cite{boson3,app51,boson5,boson6,boson7,boson8,kroha1,kroha2} and fermionic~\cite{fermion1,fermion2,fermion3,fermion4} systems. For bosons in a double-well potential, tunneling dynamics have been investigated under short- and long-range interactions~\cite{mctdhb_exact3,rhombik_epjd}, in asymmetric wells~\cite{sudip_asymmetric_dw}, in collapse–revival regimes~\cite{self_trap_r}, and in the context of self-trapping phenomena~\cite{self-trap1,self-trap2,self-trap3}. The influence of transverse degrees of freedom during tunneling has also been analyzed in detail~\cite{anal_2020_dw}. Despite these advances, relatively little attention has been given to BJJ dynamics under rotation, either at the mean-field or many-body level. 

In this work, we investigate the tunneling dynamics of interacting bosons in a two-dimensional double-well potential under rotation. We find that the tunneling dynamics are intrinsically sensitive to rotation and show that the rotation frequency can be inferred from measurable observables, including the survival probability and the expectation values of momentum and angular momentum. Our mean-field and many-body analyses show that rotation significantly modifies the tunneling behavior. The two-dimensional double-well potential is designed such that tunneling is allowed along one axis, while the transverse direction is harmonically confined. 
We first consider the configuration where the rotation axis, oriented along the z-direction, passes through the center of the double well. We show that the tunneling period, the amplitude of the transverse-momentum oscillations, and the angular momentum each exhibit a characteristic dependence on the rotation frequency. This dependence enables the rotation frequency to be determined directly from those observables. 
We then study an off-centered configuration, where the double well is shifted away from the rotation axis. In this case, the two wells experience the rotation differently, producing an asymmetry in the tunneling dynamics that increases with both the displacement and the rotation frequency. In this regime, the measured quantities allow us to infer not only the rotation frequency but also the distance of the double well from the rotation axis. Furthermore, we show that the orientation of the off-centered double well influences the tunneling dynamics. This orientation can also be inferred from the corresponding observables. 
Extending the analysis to the many-body regime shows that fragmentation is strongly affected by both the rotation frequency and the position of the double well. This observable thus offers an additional measure of inferring rotation frequencies, particularly in the slow-rotation regime. 
Finally, as an independent analysis, we consider a time-dependent rotation in the laboratory frame, where the trap is initially at rest and then gradually set into rotation. This contrasts with the previous analyses performed in the rotating frame, where the system is studied under a fixed rotation frequency. We examine the resulting tunneling dynamics for different switching times to reach the final trap rotation frequency. For sufficiently long switching times, when the system has time to adjust to the rotation, we also show how the rotation frequency can be inferred from the observed dynamics. 

The structure of the paper is as follows. In Section~\ref{system}, we introduce the system and outline the numerical method employed. Section~\ref{results} presents our mean-field results for three scenarios: (i) when the double well is at the origin, (ii) when the double well is displaced from the origin, and (iii) the orientation-dependent tunneling dynamics in the off-centered double well. For each configuration, we propose methods to infer the rotation frequency from measurable observables. In all scenarios, the rotation axis threads the origin. Many-body effects are analyzed in Section~\ref{manybody}, where fragmentation is shown to provide an additional rotation-sensitive observable. Section~\ref{labframe} presents an independent analysis of the tunneling dynamics under time-dependent rotation in the laboratory frame, where the trap itself is gradually set into motion. Section~\ref{conclusion} summarizes our findings and presents concluding remarks. Appendix~\ref{apndx} provides the analytical derivation showing how an off-centered double well can be mapped to the centered configuration. Additional many-body calculations and convergence analyses are included in the supplementary material.

%\section{Model and formalism}\label{system}
\section{System and setup}\label{system}

The tunneling dynamics of the interacting bosons in a two-dimensional double-well potential under rotation are studied by solving the many-body  Schr\"odinger equation $  i\frac{\partial \Psi}{\partial t} = \hat{H} \Psi$.
The Hamiltonian of the $N$-bosons system in the rotating frame is written as
\begin{equation} \label{hamiltonian}
\hat{H} = \sum_{i=1}^{N} \hat{h}(\mathbf{r}_i) + \sum_{i<j} \hat{W}(\mathbf{r}_i - \mathbf{r}_j) -\Omega \hat{L}_z,
\end{equation}
where the single-particle Hamiltonian is defined by $\hat{h}(\mathbf{r}_i) = -\frac{1}{2} \nabla_i^2 + V(\mathbf{r}_i)$, with $\mathbf{r_i} = (x_i, y_i)$ denoting the position in the two spatial dimensions.  $V(\mathbf{r})$ denotes the double-well potential, defined as

\begin{equation}\label{potential}
    V(x,y)= 
\begin{cases}
    \frac{1}{2}(x+2)^2+\frac{1}{2}y^2,& \text{if } x < -\frac{1}{2}\\
    \frac{3}{2}(1-x^2)+\frac{1}{2}y^2,& \text{if } -\frac{1}{2}\leq x \leq \frac{1}{2}\\
    \frac{1}{2}(x -2)^2+\frac{1}{2}y^2,& \text{if } x > \frac{1}{2}.
\end{cases}   
\end{equation}
This potential consists of two wells separated by a central barrier of finite height, forming a symmetric double-well configuration along the x-direction. The schematic diagram of the double-well potential is shown in Fig.~\ref{fig1}(a). The interaction between particles is modeled by finite-range Gaussian potential of the form 
$\hat{W}(\mathbf{r}_i - \mathbf{r}_j) = \frac{\lambda_0}{2\pi \sigma^2} \exp\left[-\frac{(\mathbf{r}_i - \mathbf{r}_j)^2}{2\sigma^2}\right]$,
where $\lambda_0$ denotes the interaction strength and $\sigma$ is fixed to $0.25$~\cite{sudip_asymmetric_dw,anal_2020_dw,beinke_rotation1}. The interaction parameter is defined as $\Lambda = \lambda_0 (N - 1) = 0.2$. 
It is important to note that, in the mean-field framework, both ground-state and dynamical properties depend solely on the interaction parameter  $\Lambda$. Consequently, the particle number $N$ can be chosen arbitrarily for a given fixed $\Lambda$.  The third term in the Hamiltonian accounts for rotation, characterized by the time-independent angular frequency $\Omega$ and the total angular momentum operator along the z-axis, $\hat{L}_z$. The rotation axis is fixed along the $\hat{z}$-direction and passes through the origin.  

The tunneling dynamics of the $N$-boson system are computed using the multiconfigurational time-dependent Hartree method for bosons (MCTDHB)~\cite{MCTDHB1,MCTDHB2}. In this framework, the many-body wave function is expressed as a superposition of time-dependent permanents constructed by distributing the $N$ bosons among $M$ orthonormal, time-dependent single-particle orbitals, $\vert \Psi(t)\rangle = \sum_{n} C_{n}(t)\vert n;t\rangle$; $\vert n;t\rangle = \prod_{i=1}^{M}  \frac{ \left( b_{i}^{\dagger}(t) \right)^{n_{i}} } {\sqrt{n_{i}!}} \vert \text{vac} \rangle$. The $M$ orthonormal single-particle states  are $\phi_j (x,t) =\langle x\vert \hat{b}_j^{\dagger}(t) \vert vac \rangle$, where $ b_{j}^{\dagger}(t)$ represents the corresponding creation operator. 

The MCTDH family of methods provides a powerful and widely used framework for studying both bosonic~\cite{MCTDHB1,MCTDHB_boson2,fischer_budha,Mistakidis_ol} and fermionic systems~\cite{MCTDHB_fermion1,MCTDHB_fermion2,paolo_fermion}. It has been applied to a broad range of physical scenarios, including tunneling dynamics in double-well potentials~\cite{sudip_asymmetric_dw,rhombik_epjd,anal_2022_dw,mctdhb_exact3,Mistakidis_dw}, ground-state properties and out-of-equilibrium dynamics in optical lattices~\cite{paolo_mott,paolo_ol,rhombik_pra,Mistakidis_ol2,rhombik_pre,paolo_cz,rhombik_shubhro_pra}, quantum motion in complex and time-dependent potentials~\cite{paolo_shake_potential,rhombik_acc}, electronic and bosonic transport~\cite{Thoss1,Thoss2,Thoss3}, indistinguishable particles coupled to optical cavities~\cite{paolo_cavity,paolo_cavity2,paolo_PhysRevA.98.053620}, bosonic spinor condensates~\cite{axel_spinor}, ultracold gases under rotation~\cite{phantom_angular_axel,rhombik_scirep,paolo_rotation,sunayana_scirep,beinke_rotation1,beinke_rotation2}, non-inertial effects~\cite{rhombik_acc}, quantum metrology~\cite{fischer_Metrology} and quantum transport in structured potentials~\cite{rhombik_jcp}. For a comprehensive overview of the method and its applications, see Refs.~\cite{mctdhb_review,MCTDHB_LectNotes}.

We solve the  Schr\"odinger equation in dimensionless units. This is achieved by scaling the Hamiltonian by $\hbar^2 / (mL_0^2)$, where $m$ is the mass of the bosons and $L_0$ represents a characteristic length scale~\cite{MCTDHB2,rhombik_acc}. In this framework, lengths in units of $L_0$ and time in units of $mL_0^2 / \hbar$. In calculations, we also adopt natural units by setting $\hbar = m= 1$. In typical experiments, the characteristic size of the trapping region is on the order of a few microns~\cite{rot_exp1,rot_exp2}. Therefore, the dimensionless parameters used in our theoretical analysis can be directly connected to experimentally relevant values by selecting an appropriate physical length scale. For instance, if we set $L_0=1 \mu$m, and consider $^{87}$Rb atoms, the corresponding unit of time becomes $t_0=mL_0^2 / \hbar=1.37\times 10^{-3}$s. 
Thus, writing the dimensionless angular frequency as $\Omega_{\mathrm{dimless}} = 2\pi/T$, where $T$ is the dimensionless period, the corresponding dimensionful angular frequency is $\Omega_{\mathrm{dim}} = \frac{\Omega_{\mathrm{dimless}}}{t_0}\;\mathrm{s}^{-1}$. Following the formation of the double-well potential, we investigate the impact of rotation on the tunneling dynamics and examine the extent to which non-inertial rotational effects are reflected in both mean-field and many-body observables.

\section{Tunneling under rotation}\label{results}
In this section, we examine the tunneling dynamics in the double-well potential under the influence of rotation. The tunneling process is strongly affected by both the rotation frequency and the position of the double well relative to the rotation axis. By analyzing the changes, we aim to identify characteristic dynamical signatures that can be used to infer the rotation frequency.

We consider a two-dimensional double-well potential in the x-y plane, with the system rotating about the z-axis and the rotation axis passing through the origin. The double well is along the x-direction, while harmonic confinement is applied along the transverse direction; see Fig.~\ref{fig1}(a). The tunneling dynamics are sensitive to the position of the double well relative to the rotation axis. When the rotation axis passes through the center of the double well [see Fig.~\ref{fig1}(a)], rotation gives rise to a centrifugal barrier that effectively pushes the bosons away from the rotation axis. The two minima of the double well are equidistant but on opposite sides of the rotation axis, so the double-well potential remains symmetric, but the effective barrier grows with increasing rotation frequency.
On the other hand, when the double well is displaced from the rotation axis [see Fig.~\ref{fig3}(a)], both wells lie on the same side of the axis. Because they are at different distances from the rotation axis, the centrifugal effect translates into an effective tilt, making the double well asymmetric. The analytical calculation of this effective tilt is given in Appendix~\ref{apndx}. 
Moreover, when the double well is not centered at the origin, the orientation of the double well in the plane becomes important. As tunneling can occur only along the axis of tunneling of the double well (and motion in the transverse direction is nearly frozen out by the harmonic trap), different orientations relative to the rotation axis lead to different tunneling behavior [see Fig.~\ref{fig5}(a)]. It is also worth noting that the Coriolis force has a negligible effect in our study, since it acts perpendicular to the motion of the particles, and as said, the motion in this direction is nearly frozen. In our setup, the transverse direction is tightly confined by the harmonic potential, suppressing any significant Coriolis contribution. 

First, we examine the case where the double well is centered at the origin (a symmetric configuration with respect to the rotation axis, so the orientation of the double well does not affect tunneling). Next, we analyze a configuration with an off-centered double well. Finally, we explore how the orientation of an off-centered double well affects the tunneling dynamics. Recall that the rotation axis is aligned with the $\hat{z}$ direction and passes through the origin.  We performed the numerical calculations in a two-dimensional box of size $ 16 \times 16$ (in length units) with periodic boundary conditions and discretized using $64 \times 64$ grid points. The convergence of the results with respect to the grid points is presented in the supplementary material.

\subsection{Basic scenario: Trap at the origin}
\begin{figure}
    \centering
    \includegraphics[width=0.99\linewidth]{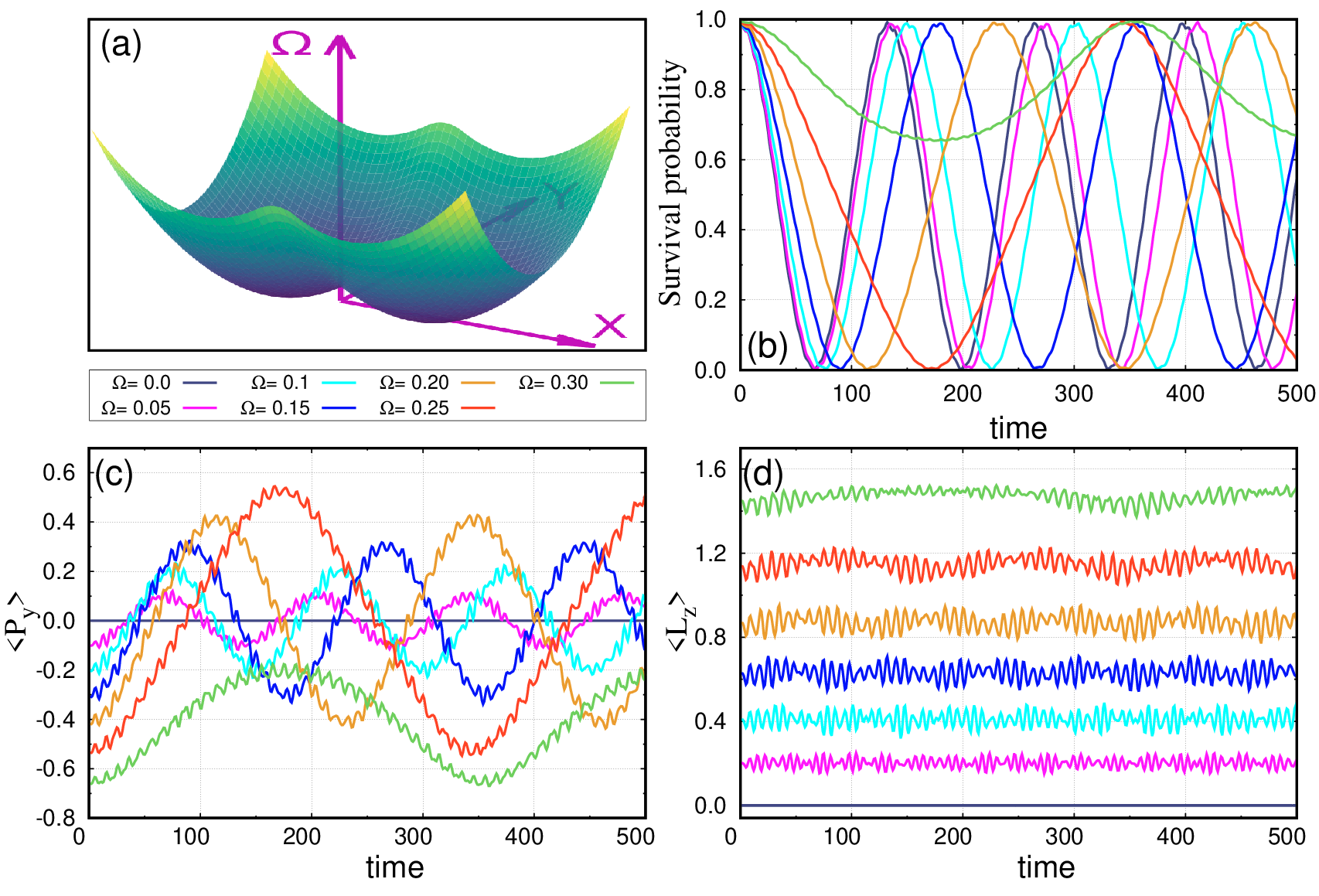}
    \caption{ \textbf{Trap centered at the rotation axis:} (a) Schematic diagram of the double-well potential centered at the rotation axis, (b) Survival probability in the left well as a function of time, (c) Expectation value of the transverse momentum per particle $\langle P_y \rangle$, and (d) Expectation value of the angular momentum per particle $\langle L_z \rangle$, shown for different rotation frequencies $\Omega$. The tunneling time period increases with $\Omega$ and eventually enters the self-trapping regime for large $\Omega$. In the full-tunneling regime, $\langle P_y \rangle$ exhibits oscillations whose period and amplitude increase with $\Omega$. $\langle L_z \rangle$ oscillates around a constant non-zero value, which also increases with $\Omega$. All results are obtained using the mean-field method, and the quantities shown are dimensionless.}
    \label{fig1}
\end{figure}

To investigate the tunneling dynamics under rotation and the associated non-inertial effects, we begin with the simplest configuration, where the double-well potential is placed at the origin; see Fig.~\ref{fig1}(a). The double well extends along the x-direction, with minima at $x=\pm2$, while harmonic confinement is applied along the y-direction. The initial state is prepared by localizing all bosons in the left well in the rotating frame with a fixed rotation frequency. This is achieved by choosing the initial potential as $V_{\text{initial}}(x,y)=\frac{1}{2}(x+2)^2+\frac{1}{2}y^2$. In the subsequent quench protocol, the double well is opened, allowing the bosons to tunnel between the two wells.

To characterize the tunneling dynamics, we observe the survival probability in the left well as a function of time and extract the corresponding tunneling period. The survival probability in the left well is obtained from the one-body density by integrating over the left half of the space. It is defined as $\mathcal{P}_L (t)= \frac{1}{N}\int_{y=-\infty}^{y=+\infty} \int_{x= -\infty}^{x=0} \rho(x,y;t) dx dy$. The survival probability also gives the information about the fraction of self-trapping during tunneling. The transverse momentum per particle is defined as $\langle P_y \rangle = \frac{1}{N}\langle \hat{p}_y \rangle = \frac{1}{N}\langle \sum_{j=1}^N \hat{p}_{y_j} \rangle$ and average angular momentum per particle is $\langle L_z \rangle = \frac{1}{N}\langle \hat{l}_z \rangle = \frac{1}{N}\langle \sum_{j=1}^N \hat{l}_{z_j} \rangle$, where $\hat{p}_y$ and $\hat{l}_z$ are the single-particle transverse momentum and angular momentum operators, respectively.

Figure~\ref{fig1}(b) shows the survival probability for different rotation frequencies $\Omega$, calculated using the mean-field densities. As shown, the tunneling period increases with increasing $\Omega$. In the absence of rotation, the tunneling dynamics exhibit a period of $T=132.4$, which we refer to as $T_{\text{Rabi}}$~\cite{anal_2020_dw,rhombik_acc}. This value agrees well with the Rabi oscillation period expected in the weakly interacting limit, given by $T_{\text{Rabi}}=\frac{2 \pi}{(E_2 -E_1)}$, where $E_2$ and $E_1$ are the lowest two energy eigenvalues of the corresponding non-interacting bosons in the double-well potential. For rotation frequencies up to $\Omega\sim 0.28$, full tunneling occurs and the tunneling period increases. Beyond this value, the system enters the self-trapping regime. This transition occurs because rotation effectively increases the barrier height of the double-well potential, thereby suppressing tunneling between the wells. For instance, at $\Omega = 0.3$, the centrifugal effect sufficiently increases the barrier to drive the system into self-trapping. The many-body survival probabilities are presented in Sec.~\ref{manybody}. In contrast to the mean-field results, the many-body survival probabilities exhibit damped oscillations due to condensate depletion during the time evolution. 

In the presence of rotation, the transverse momentum shows an oscillatory behavior, consistent with the earlier observations~\cite{sunayana_momentum}. This indicates that rotation effectively induces a double-well-like structure in momentum space along the transverse direction, a phenomena that is investigated in~\cite{sunayana_momentum}. In Fig.~\ref{fig1}(c), we show the expectation value of the momentum per particle along the y-direction. Without rotation, $\langle P_y \rangle$ remains zero. As the rotation frequency $\Omega$ increases, oscillations emerge, with both the amplitude and period increasing in the full-tunneling regime, i.e. $\Omega \le 0.28$. Upon entering the self-trapping regime, the oscillation amplitude decreases. 

In a rotating system, angular momentum naturally emerges as a key observable. The expectation value of the z-component of angular momentum per particle, $\langle L_z \rangle$, fluctuates during the tunneling dynamics, and these fluctuations are centered around an average value. Figure~\ref{fig1}(d) shows this behavior for different values of $\Omega$. Note that, for $\Omega=0$, no angular momentum is generated in the time dynamics. As the rotation frequency increases, the time-averaged value of $\langle L_z \rangle$ increases accordingly; see further analysis below. 
%%%%%%%%%%%%%%%%%%%%%%%%%%%%%%%%%%%%%%%%%%%%%

\begin{figure}
    \centering
    \includegraphics[width=0.5\linewidth]{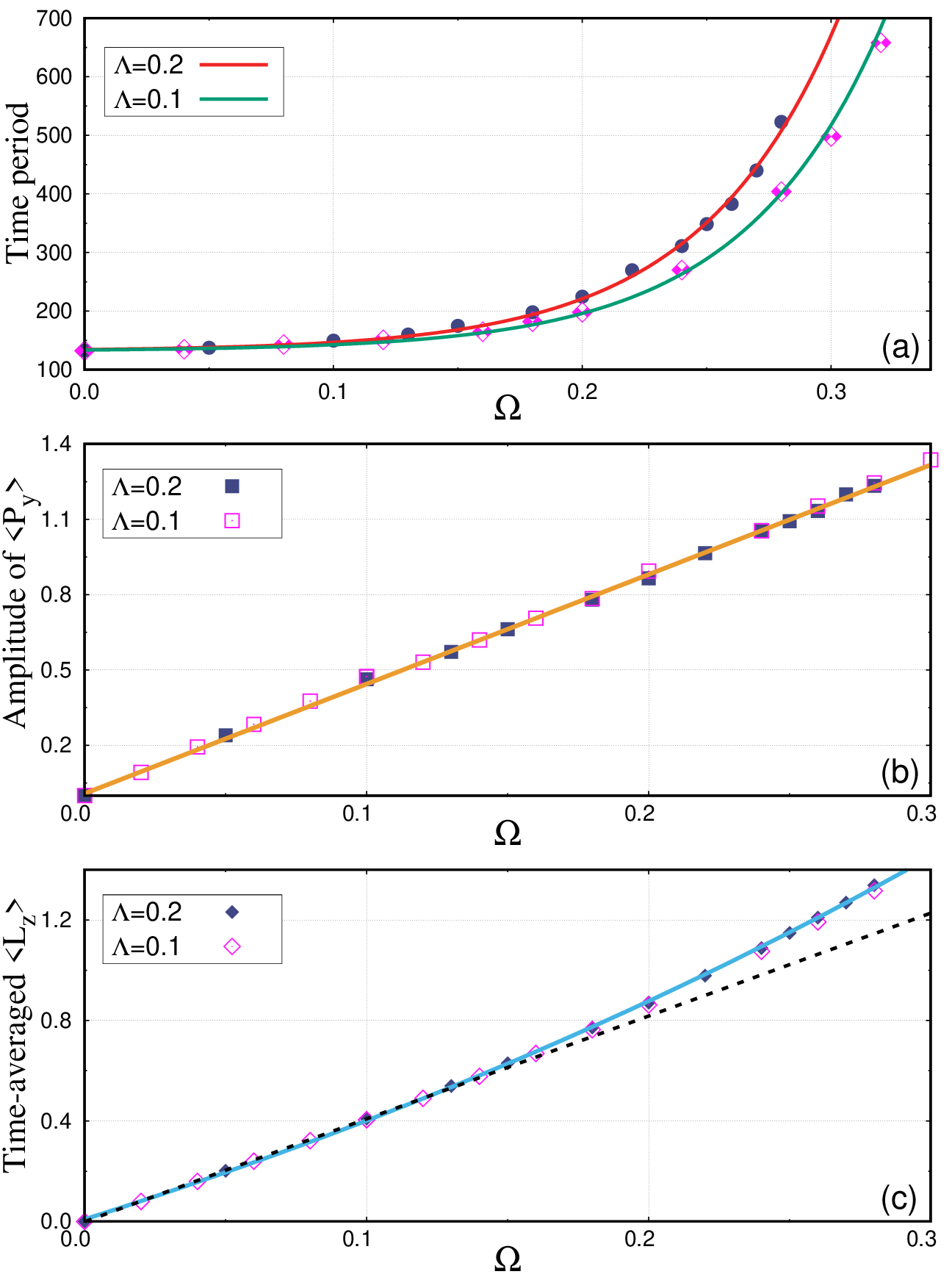}
    \caption{\textbf{Assessing rotation from observables (double well at the origin):} (a) The tunneling period as a function of rotation frequency $\Omega$, showing an exponential increase. (b) The oscillation amplitude of the expectation value of the transverse momentum $\langle P_y \rangle$ as a function of $\Omega$, showing a linear dependence. (c) The time-averaged expectation value of the angular momentum follows an exponential dependence on $\Omega$, which is also well approximated by a linear behavior in the low-$\Omega$ regime (black dotted curve). See the main text for further details. All quantities are dimensionless.}
    \label{fig2}
\end{figure}

As shown, these three observables are all sensitive to rotation. To infer the rotation frequency, we first extract the tunneling time period from the survival probability and plot it as a function of the rotation frequency $\Omega$. As found in Fig.~\ref{fig2}(a), the tunneling period increases exponentially with increasing $\Omega$. This behavior can be understood from the fact that, as $\Omega$ increases, the rotation modifies the double-well potential, effectively raising the barrier height and thereby suppressing tunneling. We additionally perform calculations for a weaker interaction strength, $\Lambda=0.1$. In both cases, the tunneling period exhibits an exponential dependence on $\Omega$. The behavior is well described by $\text{time period} \sim A \exp(17.9 \,\Omega)$, with $A=1.77$ for $\Lambda=0.1$ and $A=2.44$ for $\Lambda=0.2$. The larger prefactor for stronger interactions indicates a faster increase of the tunneling period with rotation frequency as $\Lambda$ increases. Thus, any unknown rotation frequency can be inferred solely from a measurement of the tunneling period. Note that this method is particularly sensitive in the high-rotation regime, where even a small change in $\Omega$ produces a noticeable variation in the oscillation period. In contrast, for low rotation frequencies ($\Omega \le 0.1$), the difference in the oscillation period between the rotating and non-rotating cases becomes too small to be deduced reliably using the tunneling period. Since rotation effectively enhances the barrier height and can induce self-trapping at larger $\Omega$, inferring large rotation frequencies requires starting from a sufficiently low-barrier double-well potential.

We next examine how the rotation frequency can be inferred from $\langle P_y \rangle$ and $\langle L_z \rangle$. Figure~\ref{fig2}(b) shows the amplitude of $\langle P_y \rangle$ during the dynamics for different values of $\Omega$, calculated from Fig.~\ref{fig1}(c). Within the full-tunneling regime, this amplitude increases linearly with $\Omega$, following $\langle P_y \rangle \sim 4.36\, \Omega$. Moreover, the amplitude is largely insensitive to the interaction strength, at least in the weakly interacting regime. Therefore, measuring the oscillation amplitude of $\langle P_y \rangle$ provides a direct method to infer an unknown rotation frequency, which is essentially independent of the interaction strength.  

Finally, we define the time-averaged angular momentum as $\overline{\langle L_z \rangle} = \frac{1}{T}\int_0^T \langle L_z(t) \rangle dt$.  Figure~\ref{fig2}(c) shows the time-averaged angular momentum as a function of the rotation frequency $\Omega$. The results show that the time-averaged $\langle L_z \rangle$ increases exponentially with $\Omega$. Specifically, it follows $\overline{\langle L_z \rangle} \sim 1.9\ [\exp(1.9\, \Omega) - 1]$. We also observe that, in the low-$\Omega$ regime, this behavior follows a linear dependence, as marked by the black dotted line in Fig.~\ref{fig2}(c). These behaviors are also practically independent of the interaction strength in the weakly interacting regime, as illustrated in the figure. Consequently, the rotation frequency can also be inferred from the time-averaged angular momentum.

\subsection{Tunnelling when the trap is shifted away from the rotation axis}\label{offcentred}

\begin{figure}
    \centering
    \includegraphics[width=0.99\linewidth]{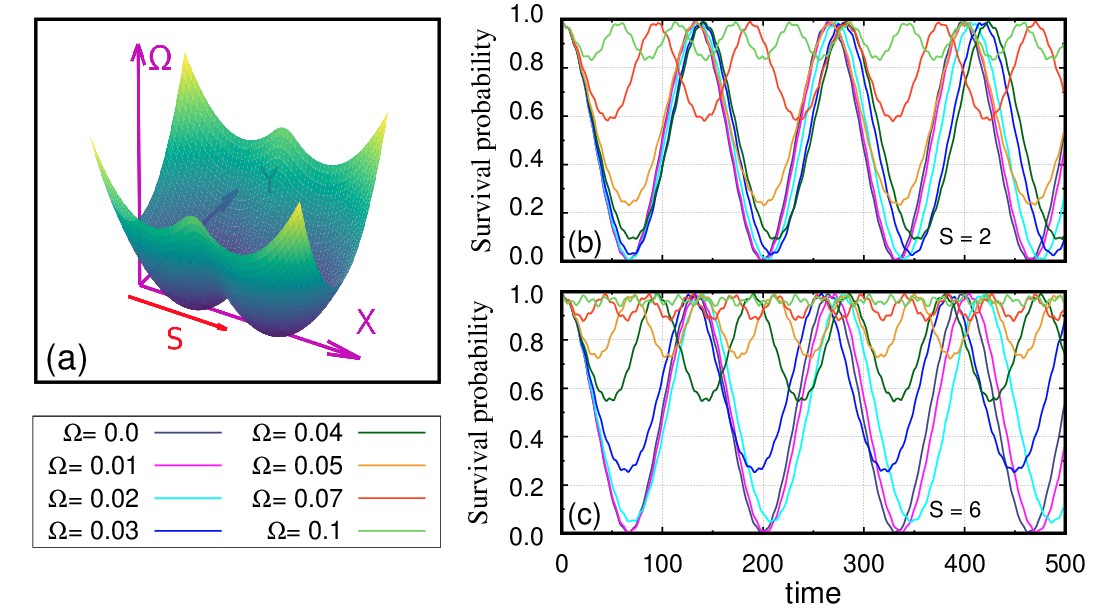}
    \caption{\textbf{Trap displaced from the rotation axis:} (a) Schematic diagram of the double-well potential displaced by a distance $S$ from the rotation axis. (b, c) Time evolution of the survival probability in the left well for different rotation frequencies $\Omega$, shown for (b) $S=2$ and (c) $S=6$. All quantities are dimensionless.}
    \label{fig3}
\end{figure}

In this section, we examine how the tunneling dynamics are affected when the double-well potential is  displaced from the center; see Fig.~\ref{fig3}(a). For a double well centered at the origin, the orientation of the well has no effect on tunneling. In contrast, once the double well is shifted away from the origin, the orientation of the trap becomes crucial. This arises for two reasons: first, rotation makes the trap asymmetric when it is displaced from the rotation axis; and second, tunneling occurs only along the double-well axis, while the bosons remain harmonically confined in the perpendicular directions. In this section, we first displace the double well along the x-axis and focus solely on the effect of its distance $S$ from the rotation axis. The influence of the orientation of an off-centered double well is discussed separately in Sec.~\ref{orientation}.

Figures~\ref{fig3}(b) and \ref{fig3}(c) show the survival probability in the left well for rotation frequencies $\Omega$, ranging from $0$ to $0.1$, for displacement values $S=2$ and $S=6$, respectively. When the double well is displaced from the origin, the potential is no longer isotropic. Shifting the double well along the x-direction effectively maps the system onto a tilted double-well potential under rotation. The analytical expression for the effective tilt is provided in Appendix~\ref{apndx}. For $S=2$ [see Fig.~\ref{fig3}(b)], in the absence of rotation, the bosons undergo complete tunneling between the wells, just like the $S=0$, $\Omega=0$ case, of course. At small rotation frequencies, full tunneling persists. As $\Omega$ increases, the system gradually enters a partial self-trapping regime and eventually transitions to strong self-trapping at higher rotation frequencies. For larger displacement [see Fig.~\ref{fig3}(c)], the same qualitative behavior is observed; however, the onset of self-trapping occurs at lower $\Omega$. Therefore, displacing the double well from the rotation center effectively introduces a tilt. This tilt increases with both the displacement $S$ and rotation frequency $\Omega$, enhancing the asymmetry between the two wells and promoting self-trapping.

\begin{figure}
    \centering
    \includegraphics[width=0.59\linewidth]{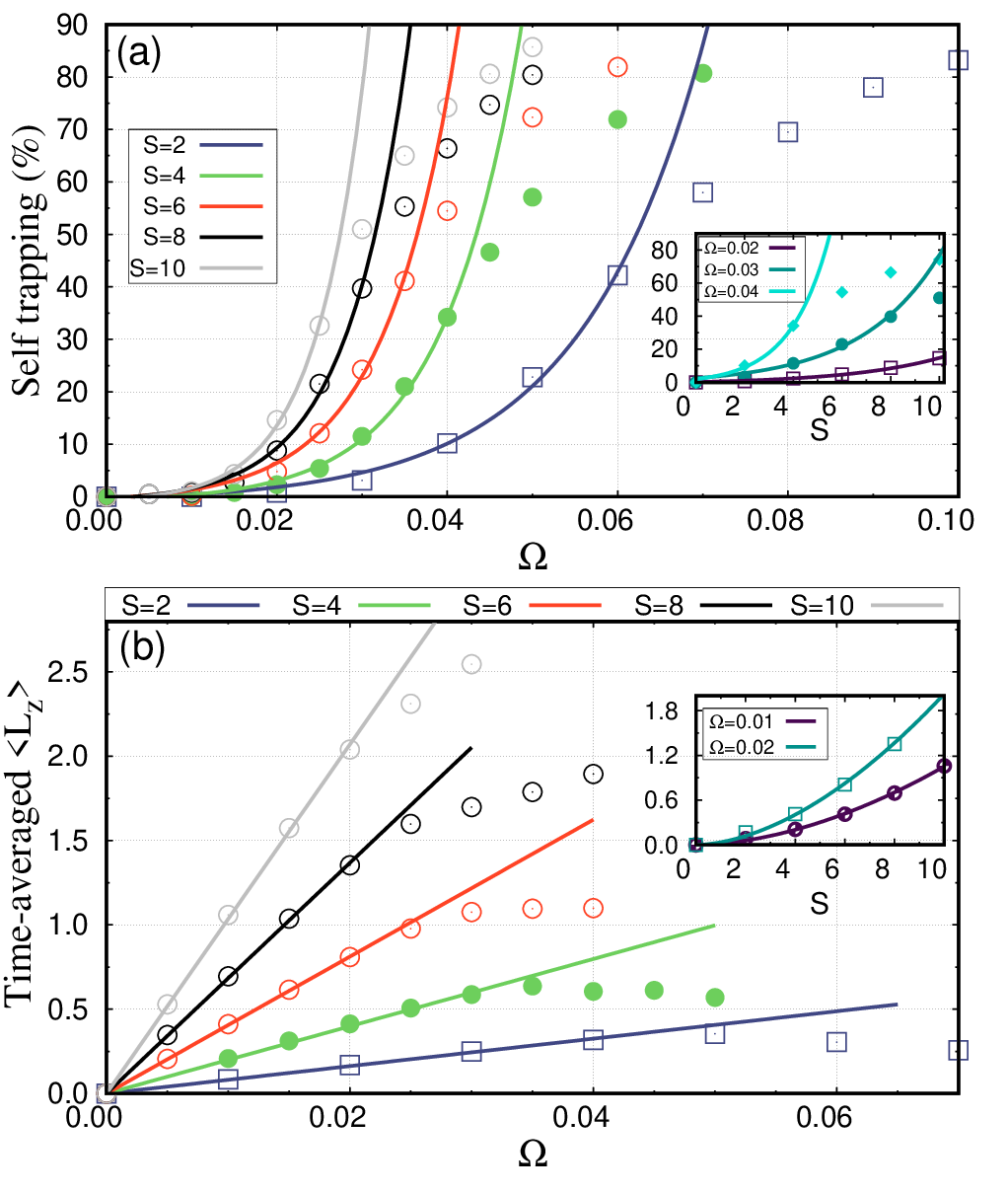}
    \caption{\textbf{Inferring slower rotation and trap displacement:} (a) Self-trapped fraction in the left well as a function of $\Omega$, extracted from the survival probability (Fig.~\ref{fig3}). The exponential fit captures the behavior well up to $\sim 40\%$ self-trapped region, beyond which the data begin to deviate from the exponential trend. Inset: Self-trapped fraction as a function of the displacement $S$, exhibiting an exponential dependence.
    (b) The time-averaged expectation value of the angular momentum per particle as a function of $\Omega$, showing a linear increase in the regime where self-trapping is not prominent. Inset: Dependence of the time-averaged angular momentum $\overline{\langle L_z \rangle}$ on $S$, showing a quadratic behavior. All quantities are dimensionless.}
    \label{fig4}
\end{figure}

We now focus on how to infer smaller rotation frequencies from the present observations. In the centered configuration, changes in the tunneling period provide a sensitive measure of rotation. When the double well is displaced, however, variations in the tunneling period become very small in the full-tunneling regime (although noticeable changes are observed in the strong self-trapping regime). In contrast, the degree of self-trapping is more sensitive to rotation in this case. Consequently, for $S \ne 0$, it is more effective to infer $\Omega$ by measuring the self-trapped fraction rather than the tunneling period. We consider the range of rotation frequencies where the tunneling gradually transitions from full tunneling to strong self-trapping. The results in Fig.~\ref{fig4}(a) show that the self-trapped fraction increases exponentially with $\Omega$ as long as it remains around or below $40\%$. The fitted exponential depends on both the displacement and the rotation frequency, allowing the rotation frequency to be extracted for a fix $S$. We also show that larger displacements lead to faster self-trapping, reflected in a larger exponential coefficient. 

Figure~\ref{fig4}(b) shows the time-averaged angular momentum per particle as a function of rotation frequency. In the nearly full-tunneling regime, the average angular momentum increases linearly with the rotation frequency. As the self-trapped fraction becomes sufficiently large, the behavior deviates from this linear trend. Furthermore, the slope becomes steeper for larger values of $S$. Thus, assessing the average angular momentum provides an additional method to infer the slower rotation frequency. 

The rotation also induces oscillations in the transverse momentum of the tunneling bosons. The amplitude of these oscillations in $\langle P_y \rangle$ depends on the magnitude of the rotation frequency. Since the present analysis focuses on small rotation frequencies well before strong self-trapping, the transverse momentum $\langle P_y \rangle$ does not provide additional useful information here and is therefore not considered further.

We now address the complementary question: if the rotation frequency can be inferred for a given $S$, can we, conversely, determine the distance of the double well from the rotation axis for a known rotation frequency by analyzing the tunneling dynamics? 
To this end, we find general relations connecting the self-trapped fraction and the time-averaged angular momentum, $\overline{\langle L_z \rangle}$, with the displacement $S$ and the rotation frequency $\Omega$. We find that the self-trapping fraction exhibits an exponential dependence described by $\text{Self-trapping\%} \sim  \exp[(11.5\,S+51)\,\Omega]$ while the time-averaged angular momentum shows a linear dependence given by $\overline{\langle L_z \rangle} \approx (S^2+4) \Omega$. The relation for the self-trapped fraction is valid when the fraction remains near or below 40\%, and the expression for $\overline{\langle L_z \rangle}$ holds in the nearly full-tunneling regime for arbitrary values of $S$ and $\Omega$. The corresponding dependence of the self-trapped fraction on $S$ is shown in the inset of Fig.~\ref{fig4}(a), while the dependence of the time-averaged angular momentum $\overline{\langle L_z \rangle}$ on $S$ is displayed in the inset of Fig.~\ref{fig4}(b). These results exhibit an exponential dependence of the self-trapped fraction on $S$ and a quadratic dependence of $\overline{\langle L_z \rangle}$ on $S$. Together, these two functional relations provide a direct means to infer either the rotation frequency or the displacement of the double-well potential from the rotation axis from the observed tunneling dynamics.

The deviation of the time-averaged angular momentum $\overline{\langle L_z \rangle}$ from a linear dependence on $\Omega$ in Fig.~\ref{fig2}(b) can be rationalized by considering the onset of partial self-trapping during the tunneling dynamics. The minima of the left and right wells are located at $x=S-2$ and $x=S+2$, respectively, such that particles localized in these wells (we ignore the width of the wave-packed for simplicity) carry angular momentum $(S-2)^2 \,\Omega$ and $(S+2)^2\, \Omega$, respectively. In the absence of self-trapping, the bosons tunnel fully between the two wells. The corresponding contributions therefore yield $\overline{\langle L_z \rangle}= \frac{1}{2}[ (S-2)^2 + (S+2)^2 ]\, \Omega = (S^2+4)\, \Omega$, consistent with the relation given above. When $P\%$ of the bosons becomes self-trapped in the left well, the equal contribution from the two wells is no longer maintained. The self-trapped fraction contributes $P\%$ times $(S-2)^2 \,\Omega$ at all times, while the remaining $(100-P)\%$ continues to tunnel between the two wells and contributes the average $\frac{1}{2}(100-P\%)[ (S-2)^2 + (S+2)^2 ]\, \Omega$. The resulting estimate for the time-averaged angular momentum is $\overline{\langle L_z \rangle} \approx \{P\%(S-2)^2 + \frac{1}{2}(100-P\%) [ (S-2)^2 + (S+2)^2 ] \}\,\Omega$, explaining the observed nonlinear dependence of $\overline{\langle L_z \rangle}$ with $\Omega$ as the fraction of self-trapping increases.

\subsection{Impact of the orientation of the double-well on tunneling}\label{orientation}
\begin{figure}
    \centering
    \includegraphics[width=0.99\linewidth]{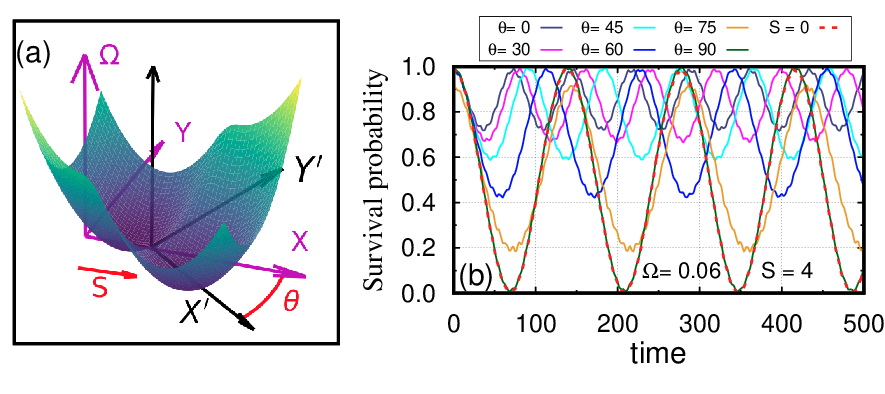}
    \caption{\textbf{Effect of displacement and orientation relative to the rotation axis:} (a) Schematic diagram of the off-center double-well potential displaced by $S$ from the rotation axis and rotated by an angle $\theta$ in the x-y plane. (b) Survival probability as a function of time for different orientations $\theta$, shown for fixed displacement $S=4$ and rotation frequency $\Omega = 0.06$. For comparison, the survival probability of the centered double well ($S=0$) is shown as a red dotted line. It coincides with the case where the double well is oriented perpendicular to the $x$-axis ($\theta = 90^\circ$). All quantities are dimensionless. }
    \label{fig5}
\end{figure}

As mentioned earlier, displacing the double well from the origin breaks the isotropy of the potential by inducing an effective tilt. In the previous subsection, we discuss how this distance from the rotation axis modifies the tunneling dynamics. Here, we examine how the orientation of the double well further influences the dynamics. To study the orientation effects, we first displace the potential by $S$ along the x-direction, such that the potential becomes $V(x-S,y)$. We then rotate the coordinates by an angle $\theta$ to define a new coordinate system $(x^{\prime},y^{\prime})$. A schematic diagram of the trap geometry and the rotation axis is shown in Fig.~\ref{fig5}(a). In this configuration, the tunneling axis is aligned with the $x^{\prime}$ direction rather than $x$, while the $y^{\prime}$ direction is harmonically confined.

For an off-centered configuration, the rotation makes the potential anisotropic and produces an effective tilt equal to $m \Omega^2Sx$ where $S$ is the displacement of the double well from the rotation axis; see Appendix~\ref{apndx}. When the tunneling axis of the double well is parallel to the direction of the centrifugal tilt [i.e., $\theta=0^{\circ}$, see Fig.~\ref{fig5}(a)], the tilt acts directly along the tunneling direction. This strongly suppresses tunneling and leads to pronounced self-trapping, as already observed in Sec.~\ref{offcentred}. In contrast, when the tunneling axis is perpendicular to the tilt ($\theta=90^{\circ}$), the centrifugal tilt is entirely along the transverse harmonic-confinement direction and has no component along the tunneling axis. As a result, the tunneling dynamics are unaffected by the off-center displacement $S$. In this case, the system behaves effectively like a centered double well. For intermediate configurations, $0^{\circ} < \theta < 90^{\circ}$, only the component of the centrifugal tilt along the $x^\prime$ axis affects the tunneling dynamics. Consequently, the effect of rotation and displacement on tunneling decreases from its maximum at $\theta=0^{\circ}$ to its minimum at $\theta=90^{\circ}$.

Fig.~\ref{fig5}(b) shows the survival probability in the left well for fixed displacement $S=4$ and rotation frequency $\Omega=0.06$, as the orientation is varied from $\theta=0^{\circ}$ to $\theta=90^{\circ}$. For reference, the centered configuration with $S=0$ and $\Omega=0.06$ is shown as well. The strongest self-trapping occurs at $\theta=0$, and the degree of self-trapping decreases as $\theta$ increases. At $\theta=90^{\circ}$, full tunneling is restored, and the dynamics coincides with that of the centered system ($S=0$, $\Omega=0.06$). This confirms that the orientation of an off-centered double well controls how strongly the rotation affects the tunneling dynamics.

\begin{figure}
    \centering
    \includegraphics[width=0.99\linewidth]{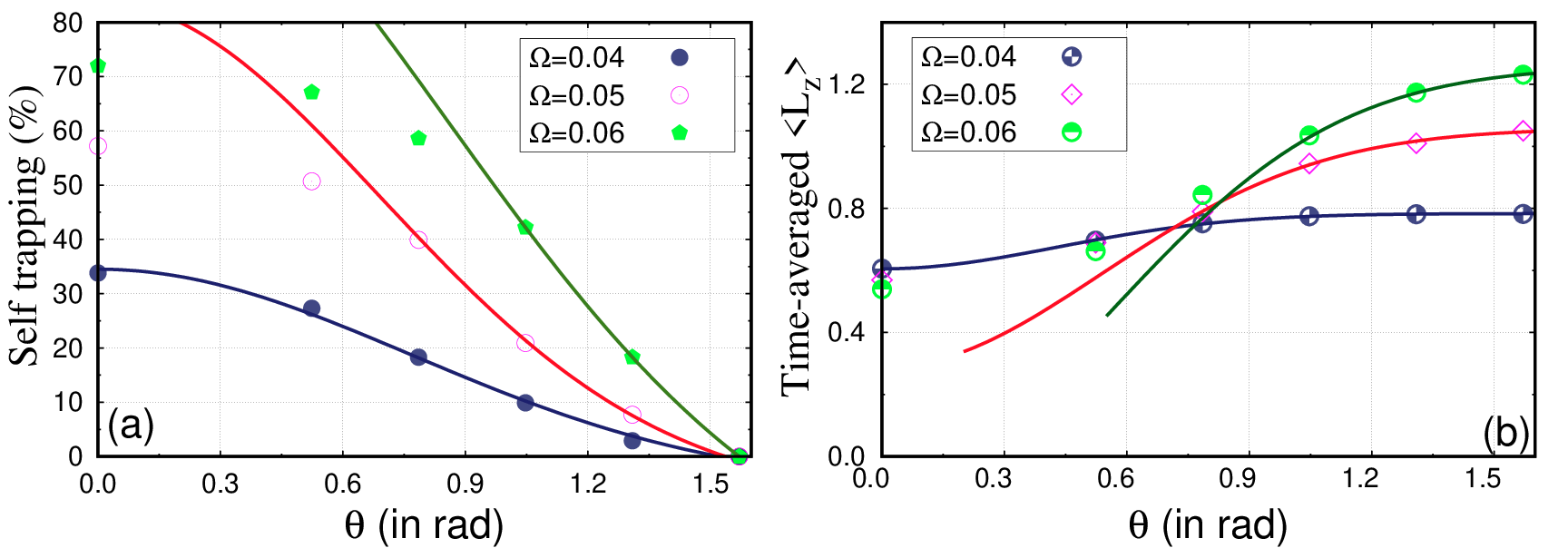}
    \caption{ \textbf{Inferring the orientation of the off-centered double well under rotation:} (a) Self-trapping percentage as a function of the orientation ($\theta$) for $S=4$, shown for three rotation frequencies $\Omega$. For self-trapping fractions $\sim 40\%$, the curves exhibit a Gaussian decrease with increasing $\theta$. The decrease rate depends on $\Omega$, allowing the rotation frequency to be inferred from the Gaussian exponent. (b) Time-averaged angular momentum per particle $\overline{\langle L_z \rangle}$ as a function of $\theta$. Below the strong self-trapping regime ($\sim 40\%$), $\overline{\langle L_z \rangle}$ shows a Gaussian increase with $\theta$. See text for more details. All quantities are dimensionless.}
    \label{fig6}
\end{figure}

We now address how the orientation $\theta$ of the double well can be inferred using the observables. For fixed displacement $S=4$, Fig.~\ref{fig6}(a) shows the percentage of self-trapping as a function of $\theta$ for three different rotation frequencies $\Omega=0.04$, $0.05$ and $0.06$. In all cases, as long as the self-trapping is around or below 40\%, the self-trapping percentage exhibits Gaussian decrease with increasing $\theta$. The characteristic widths of the Gaussian profiles depend on $\Omega$; thus, the orientation can be inferred by extracting the corresponding exponent while keeping the other parameters fixed. Another method to infer the orientation is through the expectation value of the angular-momentum. Figure~\ref{fig6}(b) displays the $\theta$ dependence of the time-averaged $\langle L_z \rangle$. As long as the self-trapping follow the Gaussian trend, $\overline{\langle L_z \rangle}$ follows a Gaussian increase with $\theta$. Together, the self-trapping percentage and the time-averaged angular momentum provide a consistent way to determine the orientation $\theta$ of an off-centered double well for a given displacement $S$ and rotation frequency $\Omega$, using the tunneling dynamics. 

\section{Many-body facets of tunneling under rotation}\label{manybody}

In this section, we analyze the many-body effects in the tunneling dynamics under rotation, focusing on how correlations influence the behavior beyond the mean-field description. To summarize and like the above, we consider three scenarios: (i) double well placed at the center, (ii) double well displaced from the center, and (iii) the off-centered double well with varying orientation. We discuss each scenario from the many-body perspective below. For the calculations, we consider $N=10$ bosons and $M=8$ orbitals. The convergence of the results is discussed in the supplementary material. 

\begin{figure}
    \centering
    \includegraphics[width=0.99\linewidth]{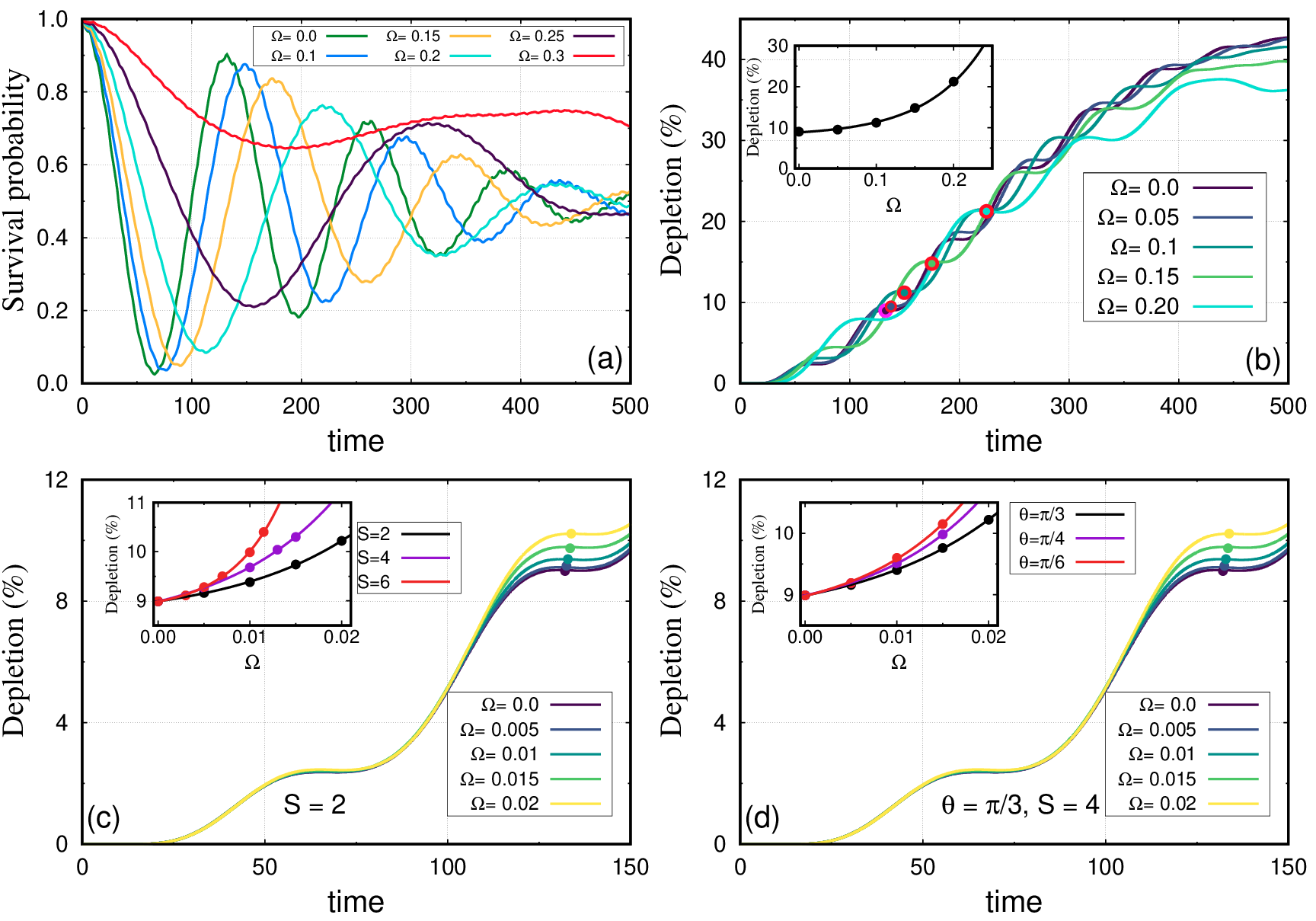}
    \caption{\textbf{Deducing rotation from many-body effects:} (a) Survival probability in the left well for the centered double well at different rotation frequencies $\Omega$, calculated from the many-body densities. The tunneling period increases with $\Omega$, and the oscillation amplitude decreases due to onset of depletion. (b) Time evolution of the many-body depletion for different $\Omega$. Inset: depletion after one tunneling period as a function of $\Omega$, showing an exponential dependence. (c) Time evolution of the many-body depletion for different $\Omega$ when the double well is off-centered with $S=2$. Inset: depletion after one tunneling period as a function of $\Omega$ for three different $S$ values, showing an exponential increase. The exponent grows with increase in $S$. (d) Depletion for an off-centered and rotated configuration ($S=4$ and $\theta=60^{\circ}$), showing similar behavior: higher rotation increases depletion, which again displays an exponential dependence after one tunneling cycle (inset). All quantities shown are dimensionless.}
    \label{fig7}
\end{figure}

When the double well is placed at the origin, the mean-field analysis exhibits that the tunneling period increases with rotation frequency $\Omega$, and eventually leading to self-trapping at larger $\Omega$. Figure~\ref{fig7}(a) shows the corresponding survival probability in the left well, calculated from the many-body densities. The results show the tunneling period increases with $\Omega$, similar to the mean-field calculations. Additionally and distinctively, the amplitude of the oscillations decreases over time. This damping of the amplitude is a hallmark of the many-body tunneling dynamics: as the system evolves, correlations build up and the condensate becomes depleted. Such decay of the oscillation amplitude in the many-body tunneling dynamics has been reported and discussed in previous studies; see, e.g., Ref.~\cite{collapse_revival,rhombik_epjd,sudip_longrange_dw,anal_2020_dw}.

Figure~\ref{fig7}(b) quantifies the increase of depletion over time for different $\Omega$. 
Depletion is defined as $D(\%) = (1-n_1) \times 100$, where $n_1$ denotes the occupation of the most populated time-dependent natural orbital. Basically, depletion quantifies the fraction of particles outside the condensate and thus serves as a useful measure of fragmentation. To investigate the effect of rotation on depletion, we measure the depletion exactly after one tunneling period for each rotation frequency, shown as solid dots. Note that the tunneling period changes with rotation frequency, so each measurement corresponds to a different time. The inset shows the depletion after one oscillation as a function of $\Omega$, which follows an exponential dependence, $\text{Depletion(\%)} \sim A \exp(13.4\, \Omega)$. Therefore, measuring depletion provides a complementary method to infer the rotation frequency. A similar mechanism has been used in the literature to infer a small linear acceleration of a moving double-well potential~\cite{rhombik_acc}. The expectation values of the transverse momentum and of the angular momentum per particle behave differently in the many-body case as well. A detailed analysis of these two quantities is provided in the supplementary material.

The survival probabilities for the other two scenarios (i.e., when the double well is off-centered and both off-centered and oriented) show the same qualitative behavior. In these cases as well, the many-body oscillation amplitude decreases over time due to the increase in depletion (not shown). In Fig.~\ref{fig7}(c), we present the depletion as a function of time for an off-centered double well displaced by $S$ ($\theta=0^{\circ}$). We present results for $S=2$, covering the dynamics up to one tunneling period and exploring rotation frequencies $\Omega$ up to $0.02$. This range of $\Omega$ is chosen because it corresponds to the regime where full tunneling occurs. Following the same procedure as before---measuring the depletion after one tunneling period for each rotation frequency---we determine the depletion after one cycle, indicated by the solid dots on the curves. We again observe that a larger $\Omega$ leads to a higher depletion. We then plot the depletion after one cycle as a function of $\Omega$ (see inset). The depletion grows exponentially with $\Omega$, indicating that depletion also serves as a sensitive indicator for detecting small rotations. Applying the same procedure for other values of $S$ (see inset), we find that the exponential dependence persists, with the exponent showing a systematic dependence on $S$. This reflects the enhanced many-body response of the system as the asymmetry increases when the potential is shifted away from the rotation axis. Based on the exponential behavior, one can determine either the rotation frequency (particularly in the slower rotation regime) or the displacement, provided the other quantity is known.

The depletion dynamics for the case where the double well is both off-centered and oriented at an angle are shown in Fig.~\ref{fig7}(d). We illustrate the tunneling behavior with one example using $S=4$ and $\theta=60^{\circ}$. In this configuration, the depletion increases with time, and after one tunneling period we measure the depletion, which is shown as the dots on the curves. The inset shows the depletion measured after one tunneling period as a function of $\Omega$, revealing an exponential dependence. We repeat the same analysis for various orientations $\theta$, and in every case, the exponential trend persists (see inset), with the exponent showing a strong dependence on $\theta$. It is worth noting that the $\theta= 0^{\circ}$ configuration corresponds exactly to the purple curve ($S=4$) in the inset of Fig.~\ref{fig7}(c). Taken together, these results indicate that depletion is also sensitive to the orientation of the off-centered double well and can therefore serve as an additional measure to infer orientation.

It is worth emphasizing that, throughout this section, we restrict our analysis to the full-tunneling regime, which necessitates considering sufficiently low values of the rotation frequency $\Omega$. Once the system enters the partial self-trapping regime, the depletion no longer follows a systematic trend, making it unsuitable as a reliable diagnostic. Although, within this range, the depletion is highly sensitive to the rotation, indicating that it provides an effective probe of slow rotation.

\section{Scenario with time-dependent rotation}\label{labframe}

In the previous sections, we analyzed tunneling dynamics in the rotating frame. The frequency of the rotation $\Omega$ is fixed, and the double-well potential is static in the rotating frame. We consider configurations in which the double well is centered on the rotation axis and where it is displaced from the rotation axis. For the off-centered case, we also explored how the orientation of the double well influences the tunneling dynamics. These analyses allow one to infer the rotation frequency and to characterize both the displacement of the double-well potential from the rotation axis and its orientation.

\begin{figure}
    \centering
    \includegraphics[width=0.6\linewidth]{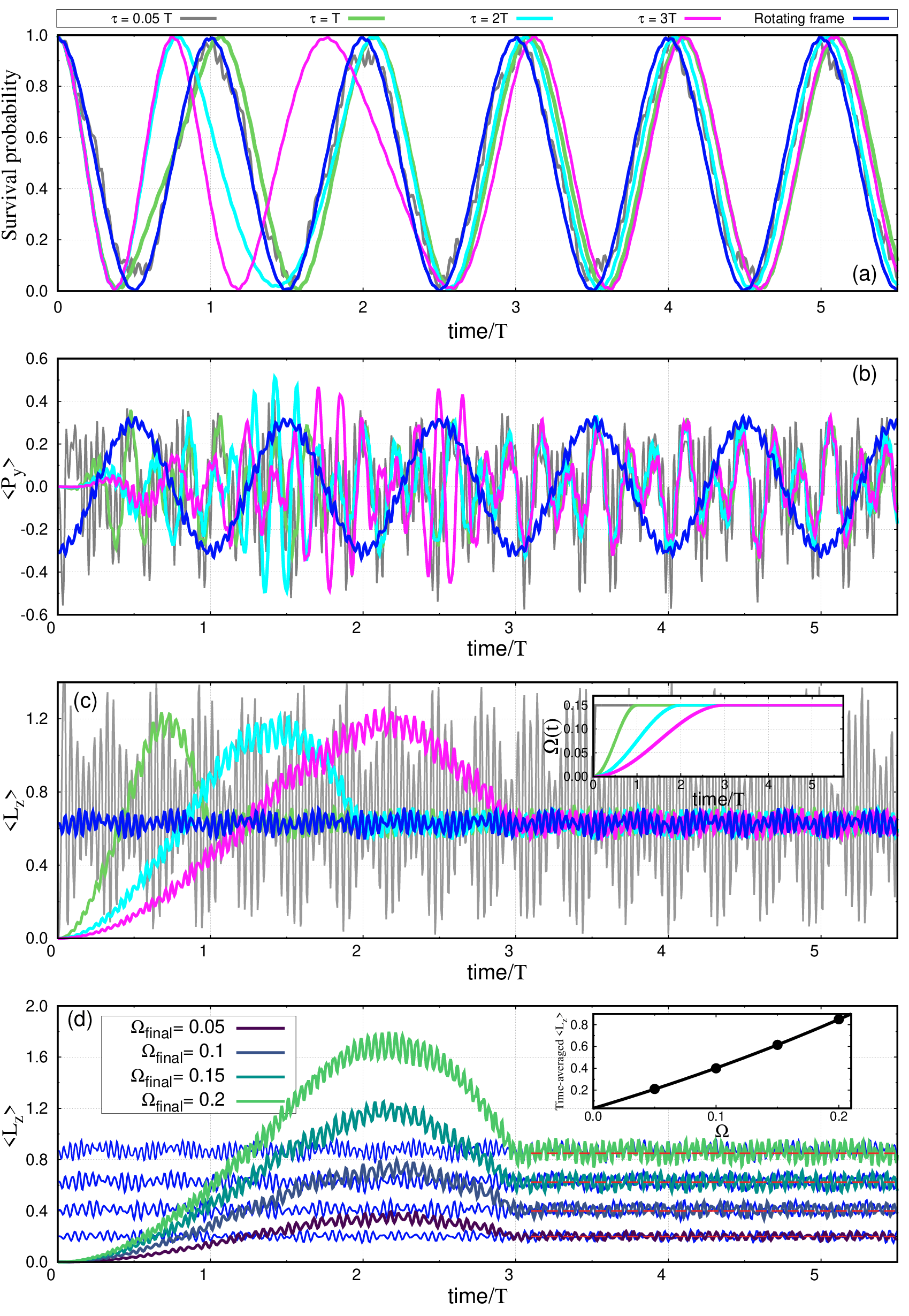}
    \caption{\textbf{Effect of time-dependent rotation on tunneling.} The bosons are initially prepared in the left well of a fixed double-well potential. The potential is then smoothly rotated for different switching times ($\tau$). (a) Survival probability: fast switching leads to irregular oscillations, while slower ramps produce smooth tunneling. (b) Transverse momentum per particle $\langle P_y \rangle$: exhibits irregular oscillations regardless of the switching time. (c) Angular momentum per particle $\langle L_z \rangle$: relaxes to a nearly steady value in the long time dynamics for all ramps except the fastest one. Results from the rotating-frame dynamics (in blue, see Fig.~\ref{fig1}) for $\Omega=0.15$ are shown for comparison. Inset: the increase of $\Omega$ over time for different switching times $\tau$. (d) Average angular momentum for different $\Omega$ at a fixed switching time of $\tau=3T$. The time-averaged angular momentum per particle, measured after the trap reaches $\Omega_{\text{final}}$ provides a way to infer the trap rotation frequency. Blue curves show the corresponding dynamics in the rotating-frame reference. The inset illustrates that the time-averaged angular momentum, $\overline{ \langle L_z \rangle}$ increases exponentially with $\Omega$. A similar exponential dependence of $\overline{\langle L_z \rangle}$ is observed in the tunneling dynamics in the rotating frame; see Fig.~\ref{fig2}(c). See text for further details. All quantities shown are dimensionless. }
    \label{fig8}
\end{figure}

In this section, we study the tunneling dynamics of bosons initially localized in the left well of a stationary double-well potential. During the quench, the double well is opened to allow tunneling, while the potential is set into rotation. To ensure a smooth transition, the trap rotation frequency is ramped up gradually to its final value. The ramp is given by $\Omega(t)= \Omega_{final} \sin^2 \left(\frac{\pi}{2}\frac{t}{\tau} \right)$; $t \le \tau$, where $\tau$ is the switching time that determines the duration required to reach the target rotation frequency. The inset of Fig.~\ref{fig8}(c) shows the time dependence of $\Omega(t)$ for different switching times. To illustrate the effect of the switching time on tunneling dynamics, we fix $\Omega_{final}=0.15$ and consider several values of $\tau$. From the rotating-frame analysis, the tunneling period at this rotation frequency is $T \approx 175$. We express the switching time in units of $T$ for convenience, and consider four cases: a very fast switch, $\tau = 0.05T$, and slower switches with $\tau = T$, $2T$, and $3T$.

Figure~\ref{fig8}(a) shows the survival probability in the left well for these switching times. For the very fast ramp ($\tau=0.05T$), the condensate experiences an abrupt change, leading to a sudden impulse, and therefore it exhibits irregular oscillations that do not reach full tunneling. For longer switching times, the system has sufficient time to adjust to the rotation, and the resulting oscillations become smooth and regular. However, the oscillations at later times still differ between different switching times. For reference, we show the survival probability for $\Omega=0.15$ (blue) in the rotating frame. Although this case is not directly comparable here, because the initial states are different, it nevertheless provides a useful qualitative reference for understanding the overall behavior. 

Figure~\ref{fig8}(b) shows the average value of the momentum operator in the y-direction per particle over time. When the double well starts to rotate, a transverse momentum develops, which is reflected in the time dynamics of $\langle P_y \rangle$. However, the nature of the graphs is not as smooth as in the rotating case of the reference frame (blue). This is because, in the present case, the system starts from a non-rotating initial state, whereas in the rotating frame, the initial state is already prepared with the rotation $\Omega$. As a result, the momentum response in the case of time-dependent rotation switching shows additional features.

Figure~\ref{fig8}(c) shows the expectation value of the angular momentum per particle $\langle L_z \rangle$. Except for the very fast switching case ($\tau = 0.05T$), the angular momentum in all other cases eventually relaxes to a steady value after the trap reaches its final rotation frequency. In the rotating frame, the angular momentum stays nearly constant from the beginning, with only small fluctuations (blue), as expected since the system is described in the rotating frame. It is also observed that as the rotation frequency $\Omega$ increases from zero to its final value, the average angular momentum per particle $\langle L_z \rangle$ initially rises and then decreases before settling to a steady value. This can be understood as follows: when the trap starts to rotate, the system begins to acquire angular momentum, causing $\langle L_z \rangle$ to grow. As the rotation becomes faster, the energy levels of the system continuously change their shape. Once $\Omega$ reaches its final value, the Hamiltonian becomes effectively time-independent, so no further excitations are generated. Some of the angular momentum gained earlier is then redistributed among different modes, which leads to a partial drop in $\langle L_z \rangle$. Eventually, $\langle L_z \rangle$ settles into a nearly constant value with small fluctuations. In short, the observed rise–fall–plateau behavior reflects how the system adjusts to the rotational excitations during tunneling. 

The behavior shown in Fig.~\ref{fig8}(c) motivates us to infer the trap rotation from $\langle L_z \rangle$. We observe that if the switching time is sufficiently long, once the trap reaches its final rotation frequency, the average angular momentum oscillates around a nearly constant value. In this regime, the time-averaged angular momentum can be calculated as  $\overline{\langle L_z \rangle} = \frac{1}{(T_f-\tau)}\int_{\tau}^{T_f} \langle L_z(t) \rangle dt$. In Fig.~\ref{fig8}(d), we plot $\langle L_z \rangle$ for different $\Omega_{\text{final}}$ values while keeping the switching time fixed at large value of $\tau = 3T$. Using this data, we compute $\overline{\langle L_z \rangle}$ (red dotted line) and present it as a function of $\Omega$ (see inset). Our results show that $\overline{\langle L_z \rangle}$ increases exponentially with the rotation frequency, following $\overline{\langle L_z \rangle} \sim 1.9\,[\exp(1.9\,\Omega)-1]$. This behavior is similar to that of the time-average angular momentum in the rotating-frame case for a centered double-well potential, see Fig.~\ref{fig2}(c).
Therefore, the trap rotation frequency, $\Omega_{\text{final}}$ can be inferred when the switching time is not too fast. 

\section{Conclusions and outlook}\label{conclusion}

We have investigated how rotation influences the tunneling dynamics of interacting bosons in a two-dimensional double-well potential. Our analysis shows that the non-inertial effects that arise due to rotation strongly control the tunneling  dynamics, and the changes in observables can be used to infer the rotation frequency. In this study, the rotation axis is chosen along the $\hat{z}$-axis passing through the origin, and the double-well potential lies in the x-y plane. For the double well placed symmetrically at the origin, the tunneling period increases exponentially with the rotation frequency, enabling a direct evaluation of the rotation frequency from it. The other observables, such as the expectation values of the transverse momentum and the time-averaged angular momentum, are also predictive of inferring the rotation frequency.

When the double well is shifted from the center, the centrifugal effects for the two wells are different. This asymmetry introduces an effective tilt in the potential, and we provided an analytical expression showing that an off-centered double well can be mapped to a centrally located double well with a rotation-induced tilt. The magnitude of this tilt is determined by the rotation frequency and the distance of the double well from the rotation axis. As a result, the tunneling dynamics differ significantly from the symmetric, centered case. In the off-centered configuration, two observables become especially informative: the self-trapping fraction and the time-averaged angular momentum. Both quantities are sensitive to the rotation frequency and to the distance of the double well from the rotation axis. We analyze how each observable varies with these parameters and show that their dependence allows one to infer both the rotation frequency and the displacement of the potential from the rotation axis. We also showed that, once the double well is displaced, the tunneling dynamics are strongly influenced by the orientation of the trap. The same two observables are used to assess the  orientation of the potential, making the system fully characterizable through measurable dynamical signatures.  

Extending the analysis to the many-body regime, we show that the depletion dynamics are strongly affected by the rotation frequency, displacement, and orientation of the double-well potential. We discuss the many-body dynamics for the three cases and offer many-body effects that provide an additional, independent means of assessing the rotation frequency, specifically in the slow rotation regime. 

Finally, we investigate the tunneling dynamics when the trap itself is rotating and show that the switching time from rest to the final rotation frequencies leaves clear signatures in the resulting dynamics. By analyzing the response to the time-dependent rotation frequency, we show that the final trap rotation frequency can be extracted from the resulting dynamics.
Overall, our results provide a framework showing that the rotation frequency, displacement from the origin, and orientation of the double well can be extracted from observables using both mean-field and many-body out-of-equilibrium dynamics.

\section*{Acknowledgements}
Computation time at the High-Performance Computing Center Stuttgart (HLRS) and High Performance Computing system Hive of the Faculty of Natural Sciences at University of Haifa are gratefully acknowledged.

\appendix
\section{Rotation-induced tilt of an off-centered double well}\label{apndx}

In the Appendix, we present the theoretical derivation showing that when a double well is displaced from the rotation axis by a distance $S$, the effect of rotation can be mapped to that of a tilted double well located at the center.
This tilt arises from centrifugal effects, which induce an effective bias between the wells, as illustrated below. 

We solve the Schr\"odinger equation $\hat{H}\Psi(\{x_i,y_i\},t) = i\frac{\partial}{\partial t} \Psi (\{x_i,y_i\},t)$, where the Hamiltonian of the rotating system when the double-well is displaced by $S$ distance along the x-axis is given by
\begin{equation}
    \hat{H} = \sum_{j=1}^N\left[\frac{\hat{p}_{x_j}^2}{2m}+\frac{\hat{p}_{y_j}^2}{2m}+V(x_j - S,y_j) -\Omega \hat{l}_{z,j}\right] + \sum_{i < j}^N\hat{W}(\mathbf{r}_i - \mathbf{r}_j). 
\end{equation}
The total angular momentum operator is $\hat{L}_{z}=\sum_{j=1}^N\hat{l}_{z,j} = \sum_{j=1}^N \left( \hat{x}_j \hat{p}_{y_j} -  \hat{y}_j \hat{p}_{x_j} \right)$. The initial state, $\Psi_0= \Psi(\{x_i,y_i\},t=0)$, orresponds to all bosons occupying the left well of a double-well potential displaced by a distance $S$. 
The two-dimensional double-well potential defined in Eq.~\ref{potential} is displaced by a distance $S$ along the x direction from the rotation axis, resulting in the shifted potential $V(x_j-S,y_j)$. The term $\hat{W}(\mathbf{r}_i - \mathbf{r}_j)$ describes the inter-particle interaction and depends only on the relative separation between two bosons. All together we start from the Schr\"odinger equation given by
 \begin{equation}\label{eq2}
  \begin{split}
    \left( \sum_{j=1}^N \left[\frac{\hat{p}_{x_j}^2}{2m}+\frac{\hat{p}_{y_j}^2}{2m}+V(x_j - S,y_j) -\Omega \hat{l}_{z,j} \right]+ \sum_{i < j}^N\hat{W}(\mathbf{r}_i - \mathbf{r}_j) \right) \Psi (\{x_j,y_j\},t) \\ = i\frac{\partial }{\partial t} \Psi (\{x_j,y_j\},t).
 \end{split}
\end{equation}
Let us introduce a transformed wave function $\Psi (\{x_j,y_j\},t) = e^{-i\sum_l S \hat{p}_{x_l}} \Phi (\{x_j,y_j\},t)$. Substituting this expression into Eq.~\ref{eq2}, we obtain

 \begin{equation}
 \begin{split}
\left(  \sum_{j=1}^N \left[ \frac{\hat{p}_{x_j}^2}{2m}+\frac{\hat{p}_{y_j}^2}{2m}+V(x_j - S,y_j) -\Omega \hat{l}_{z,j}\right] + \sum_{i < j}^N\hat{W}(\mathbf{r}_i - \mathbf{r}_j) \right) e^{-i\sum_l S \hat{p}_{x_l}} \Phi (\{x_j,y_j\},t) \\ =  i\frac{\partial }{\partial t} \left( e^{-i\sum_l S \hat{p}_{x_l}} \Phi (\{x_j,y_j\},t) \right).
    \end{split}
 \end{equation}
Note that the interaction term does not change or play any role in this transformation. It depends only on the
relative distance between the bosons. 
 \begin{equation}\label{eq4}
 \begin{split}
\left( \sum_{j=1}^N \left[\frac{\hat{p}_{x_j}^2}{2m}+\frac{\hat{p}_{y_j}^2}{2m}+V(x_j,y_j) -\Omega \hat{l}_{z,j} +\Omega S \hat{p}_{y_j} \right]+ \sum_{i < j}^N\hat{W}(\mathbf{r}_i - \mathbf{r}_j)\right)  \Phi (\{x_j,y_j\},t) \\ = i\frac{\partial }{\partial t} \left(  \Phi (\{x_j,y_j\},t) \right).
    \end{split}
 \end{equation}

Rearranging Eq.~\ref{eq4}, we obtain
\begin{equation}\label{eq5}
\begin{split}
    \left( \sum_{j=1}^N\left[ \frac{\hat{p}_{x_j}^2}{2m}+\frac{(\hat{p}_{y_j} + m \Omega s)^2}{2m} - \frac{m \Omega^2 S^2}{2}+V(x_j,y_j) - \Omega \hat{l}_{z,j} \right]+ \sum_{i < j}^N\hat{W}(\mathbf{r}_i - \mathbf{r}_j)\right) \Phi (\{x_j,y_j\},t) \\= i\frac{\partial}{\partial t}\Phi (\{x_j,y_j\},t).
\end{split}
\end{equation}

This transformation shifts the potential to the center but introduces a constant term with the momentum operator $\hat{p}_y$. Let us do another transformation to get rid of the $m\Omega S$ term from $\hat{p}_y$. The transformation follows $\Phi (\{x_j,y_j\},t) = e^{-i \sum_l m \Omega S y_l} \xi (\{x_j,y_j\},t)$. Substituting it into Eq.~\ref{eq5} we get

\begin{equation}
\begin{split}
    \left( \sum_{j=1}^N\left[ \frac{\hat{p}_{x_j}^2}{2m}+\frac{(\hat{p}_{y_j} + m \Omega s)^2}{2m} - \frac{m \Omega^2 S^2}{2}+V(x_j,y_j) - \Omega \hat{l}_{z,j} \right]+ \sum_{i < j}^N\hat{W}(\mathbf{r}_i - \mathbf{r}_j)\right) \\ e^{-i \sum_l m \Omega S y_l} \xi (\{x_j,y_j\},t)  = i\frac{\partial}{\partial t}e^{-i \sum_l m \Omega S y_l} \xi (\{x_j,y_j\},t),    
\end{split}
\end{equation}

\begin{equation}
\begin{split}
    \left( \sum_{j=1}^N\left[ \frac{\hat{p}_{x_j}^2}{2m}+\frac{\hat{p}_{y_j}^2}{2m} +V(x_j,y_j) - \Omega \hat{l}_{z,j} -m \Omega^2Sx_j - \frac{m \Omega^2 S^2}{2} \right] + \sum_{i < j}^N\hat{W}(\mathbf{r}_i - \mathbf{r}_j) \right) \\ \xi (\{x_j,y_j\},t)   = i\frac{\partial}{\partial t} \xi (\{x_j,y_j\},t).    
\end{split}
\end{equation} 
Finally, we obtain the full Schr\"odinger equation after applying two transformations:
 \begin{equation}\label{eq8}
 \begin{split}
    \left( \sum_{j=1}^N \left[\frac{\hat{p}_{x_j}^2}{2m}+\frac{\hat{p}_{y_j}^2}{2m}+ \{V(x_j,y_j)-m \Omega^2Sx_j\} -\Omega \hat{l}_{z,j} - \frac{m \Omega^2 S^2}{2} \right] + \sum_{i < j}^N\hat{W}(\mathbf{r}_i - \mathbf{r}_j) \right) &\\ \xi (\{x_j,y_j\},t) = i\frac{\partial }{\partial t} \xi (\{x_j,y_j\},t).
\end{split}
 \end{equation}
The Schr\"odinger equation acquires a modified potential of the form $\{V(x_i,y_i)-m \Omega^2Sx_i\}$, where $V(x_i,y_i)$ is the double-well potential centered on the rotation axis, and the term $m \Omega^2 S$ introduces an effective tilt. Thus, a double well that is displaced from the rotation axis can be mapped onto an equivalent double well at the center, but with a tilt whose strength is proportional to the displacement $S$ and to the square of the rotation frequency $\Omega$. The initial wave function of the off-centered double well and that of the centered double well are related through $ \xi(\mathbf{r},t=0) = e^{+i \sum_j m \Omega S y_j}e^{+i \sum_k S \hat{p}_{x_k}}   \Psi_0$. 
The constant term $\frac{m \Omega^2 S^2}{2}$ corresponds to a constant energy shift and can be removed by an additional phase transformation.

\bibliographystyle{apsrev4-2-title} 
\bibliography{ref}

@article{rhombik_scirep,
author={Roy, Rhombik
and Dutta, Sunayana
and Alon, Ofir E.},
title={Rotation quenches in trapped bosonic systems},
journal={Sci. Rep.},
year={2025},
month={Jul},
day={26},
volume={15},
number={1},
pages={27193},
issn={2045-2322},
doi={10.1038/s41598-025-07144-w},
urrl={https://doi.org/10.1038/s41598-025-07144-w}
}

@article{rhombik_shubhro_pra,
  title = {One-dimensional quench dynamics in an optical lattice: {Sine-Gordon} and {Bose-Hubbard} descriptions},
  author = {Roy, Subhrajyoti and Roy, Rhombik and Trombettoni, Andrea and Chakrabarti, Barnali and Gammal, Arnaldo},
  journal = {Phys. Rev. A},
  volume = {111},
  issue = {1},
  pages = {013315},
  numpages = {12},
  year = {2025},
  month = {Jan},
  publisher = {American Physical Society},
  doi = {10.1103/PhysRevA.111.013315},
  url = {https://link.aps.org/doi/10.1103/PhysRevA.111.013315}
}

@article{rhombik_jcp,
    author = {Roy, Rhombik and Alon, Ofir E.},
    title = {Dynamics and transport of {Bose–Einstein} condensates in bent potentials},
    journal = {J. Chem. Phys.},
    volume = {163},
    number = {22},
    pages = {224306},
    year = {2025},
    month = {12},
    issn = {0021-9606},
    doi = {10.1063/5.0301304},
    url = {https://doi.org/10.1063/5.0301304}
}

@article{rhombik_acc,
  title = {Assessing small accelerations using a bosonic {Josephson} junction},
  author = {Roy, Rhombik and Alon, Ofir E.},
  journal = {Phys. Rev. A},
  volume = {111},
  issue = {4},
  pages = {043307},
  numpages = {14},
  year = {2025},
  month = {Apr},
  publisher = {American Physical Society},
  doi = {10.1103/PhysRevA.111.043307},
  urrl = {https://link.aps.org/doi/10.1103/PhysRevA.111.043307}
}

@article{rhombik_pra,
  title = {Phases, many-body entropy measures, and coherence of interacting bosons in optical lattices},
  author = {Roy, R. and Gammal, A. and Tsatsos, M. C. and Chatterjee, B. and Chakrabarti, B. and Lode, A. U. J.},
  journal = {Phys. Rev. A},
  volume = {97},
  issue = {4},
  pages = {043625},
  numpages = {10},
  year = {2018},
  month = {Apr},
  publisher = {American Physical Society},
  doi = {10.1103/PhysRevA.97.043625},
  urrl = {https://link.aps.org/doi/10.1103/PhysRevA.97.043625}
}

@article{rhombik_epjd,
	doi = {10.1140/epjd/s10053-022-00345-2},
	urrl = {https://doi.org/10.1140/epjd/s10053-022-00345-2},
	year = 2022,
	month = {feb},
	publisher = {Springer},
	volume = {76},
	number = {24},
	pages = {215303},
	author = {Rhombik Roy and Barnali Chakrabarti and Andrea Trombettoni},
	title = {Quantum dynamics of few dipolar bosons in a double-well potential.},
	journal = {Eur. Phys. J. D}
}

@article{rhombik_pre,
  title = {Information theoretic measures for interacting bosons in optical lattice},
  author = {Roy, Rhombik and Chakrabarti, Barnali and Chavda, N. D. and Lekala, M. L.},
  journal = {Phys. Rev. E},
  volume = {107},
  issue = {2},
  pages = {024119},
  numpages = {11},
  year = {2023},
  month = {Feb},
  publisher = {American Physical Society},
  doi = {10.1103/PhysRevE.107.024119},
  urrl = {https://link.aps.org/doi/10.1103/PhysRevE.107.024119}
}

@Article{sunayana_scirep,
author={Dutta, Sunayana
and Lode, Axel U. J.
and Alon, Ofir E.},
title={Fragmentation and correlations in a rotating {Bose-Einstein} condensate undergoing breakup},
journal={Sci. Rep.},
year={2023},
month={Feb},
day={27},
volume={13},
number={1},
pages={3343},
issn={2045-2322},
doi={10.1038/s41598-023-29516-w},
urrl={https://doi.org/10.1038/s41598-023-29516-w}
}

@article{sudip_asymmetric_dw,
doi = {10.1088/1367-2630/ab4315},
urrl = {https://dx.doi.org/10.1088/1367-2630/ab4315},
year = {2019},
month = {oct},
publisher = {IOP Publishing},
volume = {21},
number = {10},
pages = {103037},
author = {Sudip Kumar Haldar and Ofir E Alon},
title = {Many-body quantum dynamics of an asymmetric bosonic {Josephson} junction},
journal = {New J. Phys.}
}

@article{sudip_longrange_dw,
title = {Impact of the range of the interaction on the quantum dynamics of a bosonic {Josephson} junction},
journal = {Chem. Phys.},
volume = {509},
pages = {72},
year = {2018},
isssn = {0301-0104},
doi = {https://doi.org/10.1016/j.chemphys.2018.01.017},
urrl = {https://www.sciencedirect.com/science/article/pii/S0301010417308881},
author = {Sudip Kumar Haldar and Ofir E. Alon}
}

@Article{anal_2020_dw,
author={Bhowmik, Anal
and Haldar, Sudip Kumar
and Alon, Ofir E.},
title={Impact of the transverse direction on the many-body tunneling dynamics in a two-dimensional bosonic {Josephson} junction},
journal={Sci. Rep.},
year={2020},
month={Dec},
day={08},
volume={10},
number={1},
pages={21476},
isssn={2045-2322},
doi={10.1038/s41598-020-78173-w},
urrl={https://doi.org/10.1038/s41598-020-78173-w}
}

@Article{anal_2022_dw,
author={Bhowmik, Anal
and Alon, Ofir E.},
title={Longitudinal and transversal resonant tunneling of interacting bosons in a two-dimensional {Josephson} junction},
journal={Sci. Rep.},
year={2022},
month={Jan},
day={12},
volume={12},
number={1},
pages={627},
isssn={2045-2322},
doi={10.1038/s41598-021-04312-6},
urrl={https://doi.org/10.1038/s41598-021-04312-6}
}

@mice{sunayana_momentum,
title={Rotation-mediated bosonic {Josephson} junctions in position and momentum spaces}, 
author={Sunayana Dutta and Ofir E. Alon},
year={2025},
eprint={2503.10153},
archivePrefix={arXiv},
primaryClass={cond-mat.quant-gas},
url={https://arxiv.org/abs/2503.10153}, 
}

@article{fetter,
  title = {Quantized superfluid vortex dynamics on cylindrical surfaces and planar annuli},
  author = {Guenther, Nils-Eric and Massignan, Pietro and Fetter, Alexander L.},
  journal = {Phys. Rev. A},
  volume = {96},
  issue = {6},
  pages = {063608},
  numpages = {14},
  year = {2017},
  month = {Dec},
  publisher = {American Physical Society},
  doi = {10.1103/PhysRevA.96.063608},
  urrl = {https://link.aps.org/doi/10.1103/PhysRevA.96.063608}
}

@Article{MCTDHB_LectNotes,
	title={{Many-body quantum dynamics with MCTDH-X}},
	author={Paolo Molignini and Sunayana Dutta and Elke Fasshauer},
	journal={SciPost Phys. Lect. Notes},
	volume={94},
	year={2025},
	publisher={SciPost},
	dooi={10.21468/SciPostPhysLectNotes.94},
	url={https://scipost.org/10.21468/SciPostPhysLectNotes.94},
}

@article{intro1,
  title = {Quantum Coherent Atomic Tunneling between Two Trapped {Bose-Einstein} Condensates},
  author = {Smerzi, A. and Fantoni, S. and Giovanazzi, S. and Shenoy, S. R.},
  journal = {Phys. Rev. Lett.},
  volume = {79},
  issue = {25},
  pages = {4950},
  numpages = {0},
  year = {1997},
  month = {Dec},
  publisher = {American Physical Society},
  doi = {10.1103/PhysRevLett.79.4950},
  urrl = {https://link.aps.org/doi/10.1103/PhysRevLett.79.4950}
}

@article{intro3,
  title = {Superconductor-insulator transition in {Josephson} junction chains by quantum {Monte Carlo} calculations},
  author = {Basko, D. M. and Pfeiffer, F. and Adamus, P. and Holzmann, M. and Hekking, F. W. J.},
  journal = {Phys. Rev. B},
  volume = {101},
  issue = {2},
  pages = {024518},
  numpages = {11},
  year = {2020},
  month = {Jan},
  publisher = {American Physical Society},
  doi = {10.1103/PhysRevB.101.024518},
  urrl = {https://link.aps.org/doi/10.1103/PhysRevB.101.024518}
}

@article{fischer_Metrology,
  title = {Self-Consistent Many-Body Metrology},
  author = {Baak, Jae-Gyun and Fischer, Uwe R.},
  journal = {Phys. Rev. Lett.},
  volume = {132},
  issue = {24},
  pages = {240803},
  numpages = {6},
  year = {2024},
  month = {Jun},
  publisher = {American Physical Society},
  doi = {10.1103/PhysRevLett.132.240803},
  urrl = {https://link.aps.org/doi/10.1103/PhysRevLett.132.240803}
}

@article{fischer_budha,
  title = {Condensate fragmentation as a sensitive measure of the quantum many-body behavior of bosons with long-range interactions},
  author = {Fischer, Uwe R. and Lode, Axel U. J. and Chatterjee, Budhaditya},
  journal = {Phys. Rev. A},
  volume = {91},
  issue = {6},
  pages = {063621},
  numpages = {8},
  year = {2015},
  month = {Jun},
  publisher = {American Physical Society},
  doi = {10.1103/PhysRevA.91.063621},
  urrl = {https://link.aps.org/doi/10.1103/PhysRevA.91.063621}
}

@article{boson3,
  title = {Crossing Over from Attractive to Repulsive Interactions in a Tunneling Bosonic {Josephson} Junction},
  author = {Spagnolli, G. and Semeghini, G. and Masi, L. and Ferioli, G. and Trenkwalder, A. and Coop, S. and Landini, M. and Pezz\`e, L. and Modugno, G. and Inguscio, M. and Smerzi, A. and Fattori, M.},
  journal = {Phys. Rev. Lett.},
  volume = {118},
  issue = {23},
  pages = {230403},
  numpages = {6},
  year = {2017},
  month = {Jun},
  publisher = {American Physical Society},
  doi = {10.1103/PhysRevLett.118.230403},
  urrl = {https://link.aps.org/doi/10.1103/PhysRevLett.118.230403}
}

@article{boson5,
  title = {Dynamics of a Tunable Superfluid Junction},
  author = {LeBlanc, L. J. and Bardon, A. B. and McKeever, J. and Extavour, M. H. T. and Jervis, D. and Thywissen, J. H. and Piazza, F. and Smerzi, A.},
  journal = {Phys. Rev. Lett.},
  volume = {106},
  issue = {2},
  pages = {025302},
  numpages = {4},
  year = {2011},
  month = {Jan},
  publisher = {American Physical Society},
  doi = {10.1103/PhysRevLett.106.025302},
  urrl = {https://link.aps.org/doi/10.1103/PhysRevLett.106.025302}
}

@Article{boson6,
author={Levy, S.
and Lahoud, E.
and Shomroni, I.
and Steinhauer, J.},
title={The a.c. and d.c. {Josephson} effects in a {Bose-Einstein} condensate},
journal={Nature},
year={2007},
month={Oct},
day={01},
volume={449},
number={7162},
pages={579},
isssn={1476-4687},
doi={10.1038/nature06186},
urrl={https://doi.org/10.1038/nature06186}
}

@article{boson7,
  title = {Direct Observation of Tunneling and Nonlinear Self-Trapping in a Single Bosonic {Josephson} Junction},
  author = {Albiez, Michael and Gati, Rudolf and F\"olling, Jonas and Hunsmann, Stefan and Cristiani, Matteo and Oberthaler, Markus K.},
  journal = {Phys. Rev. Lett.},
  volume = {95},
  issue = {1},
  pages = {010402},
  numpages = {4},
  year = {2005},
  month = {Jun},
  publisher = {American Physical Society},
  doi = {10.1103/PhysRevLett.95.010402},
  urrl = {https://link.aps.org/doi/10.1103/PhysRevLett.95.010402}
}

@Article{boson8,
author={Est{\`e}ve, J.
and Gross, C.
and Weller, A.
and Giovanazzi, S.
and Oberthaler, M. K.},
title={Squeezing and entanglement in a {Bose-Einstein} condensate},
journal={Nature},
year={2008},
month={Oct},
day={01},
volume={455},
number={7217},
pages={1216},
isssn={1476-4687},
doi={10.1038/nature07332},
urrl={https://doi.org/10.1038/nature07332}
}

@article{fermion1,
author = {W. J. Kwon  and G. Del Pace  and R. Panza  and M. Inguscio  and W. Zwerger  and M. Zaccanti  and F. Scazza  and G. Roati },
title = {Strongly correlated superfluid order parameters from dc {Josephson} supercurrents},
journal = {Science},
volume = {369},
number = {6499},
pages = {84},
year = {2020},
doi = {10.1126/science.aaz2463},
urrl = {https://www.science.org/doi/abs/10.1126/science.aaz2463},
epprint = {https://www.science.org/doi/pdf/10.1126/science.aaz2463}}

@article{fermion2,
author = {Niclas Luick  and Lennart Sobirey  and Markus Bohlen  and Vijay Pal Singh  and Ludwig Mathey  and Thomas Lompe  and Henning Moritz },
title = {An ideal {Josephson} junction in an ultracold two-dimensional {Fermi} gas},
journal = {Science},
volume = {369},
number = {6499},
pages = {89},
year = {2020},
doi = {10.1126/science.aaz2342},
urrl = {https://www.science.org/doi/abs/10.1126/science.aaz2342}}

@article{fermion3,
author = {Giacomo Valtolina  and Alessia Burchianti  and Andrea Amico  and Elettra Neri  and Klejdja Xhani  and Jorge Amin Seman  and Andrea Trombettoni  and Augusto Smerzi  and Matteo Zaccanti  and Massimo Inguscio  and Giacomo Roati },
title = {{Josephson} effect in fermionic superfluids across the {BEC-BCS} crossover},
journal = {Science},
volume = {350},
number = {6267},
pages = {1505},
year = {2015},
doi = {10.1126/science.aac9725},
urrl = {https://www.science.org/doi/abs/10.1126/science.aac9725},
epprint = {https://www.science.org/doi/pdf/10.1126/science.aac9725}}

@article{fermion4,
  title = {Connecting Dissipation and Phase Slips in a {Josephson} Junction between Fermionic Superfluids},
  author = {Burchianti, A. and Scazza, F. and Amico, A. and Valtolina, G. and Seman, J. A. and Fort, C. and Zaccanti, M. and Inguscio, M. and Roati, G.},
  journal = {Phys. Rev. Lett.},
  volume = {120},
  issue = {2},
  pages = {025302},
  numpages = {6},
  year = {2018},
  month = {Jan},
  publisher = {American Physical Society},
  doi = {10.1103/PhysRevLett.120.025302},
  urrl = {https://link.aps.org/doi/10.1103/PhysRevLett.120.025302}
}

@article{self-trap1,
doi = {10.1088/1361-6455/ace66d},
urrl = {https://dx.doi.org/10.1088/1361-6455/ace66d},
year = {2023},
month = {jul},
publisher = {IOP Publishing},
volume = {56},
number = {16},
pages = {165301},
author = {Fatkhulla Kh Abdullaev and Ravil M Galimzyanov and Akbar M Shermakhmatov},
title = {Effects of quantum fluctuations on macroscopic quantum tunneling and self-trapping of a {BEC} in a double-well trap},
journal = {J. Phys. B At. Mol. Opt. Phys.}
}

@article{self-trap2,
doi = {10.1088/1361-6455/ab2b58},
urrl = {https://dx.doi.org/10.1088/1361-6455/ab2b58},
year = {2019},
month = {jul},
publisher = {IOP Publishing},
volume = {52},
number = {15},
pages = {155301},
author = {Abhik Kumar Saha and Kingshuk Adhikary and Subhanka Mal and Krishna Rai Dastidar and Bimalendu Deb},
title = {The effects of trap-confinement and interatomic interactions on {Josephson} effects and macroscopic quantum self-trapping for a {Bose–Einstein} condensate},
journal = {J. Phys. B At. Mol. Opt. Phys.}
}

@article{self-trap3,
  title = {Classical Bifurcation at the Transition from {Rabi} to {Josephson} Dynamics},
  author = {Zibold, Tilman and Nicklas, Eike and Gross, Christian and Oberthaler, Markus K.},
  journal = {Phys. Rev. Lett.},
  volume = {105},
  issue = {20},
  pages = {204101},
  numpages = {4},
  year = {2010},
  month = {Nov},
  publisher = {American Physical Society},
  doi = {10.1103/PhysRevLett.105.204101},
  urrl = {https://link.aps.org/doi/10.1103/PhysRevLett.105.204101}
}

@article{self_trap_r,
  title = {Quantum dynamics of an atomic {Bose-Einstein} condensate in a double-well potential},
  author = {Milburn, G. J. and Corney, J. and Wright, E. M. and Walls, D. F.},
  journal = {Phys. Rev. A},
  volume = {55},
  issue = {6},
  pages = {4318},
  numpages = {0},
  year = {1997},
  month = {Jun},
  publisher = {American Physical Society},
  doi = {10.1103/PhysRevA.55.4318},
  urrl = {https://link.aps.org/doi/10.1103/PhysRevA.55.4318}
}

@article{MCTDHB2,
  title = {Role of Excited States in the Splitting of a Trapped Interacting {Bose-Einstein} Condensate by a Time-Dependent Barrier},
  author = {Streltsov, Alexej I. and Alon, Ofir E. and Cederbaum, Lorenz S.},
  journal = {Phys. Rev. Lett.},
  volume = {99},
  issue = {3},
  pages = {030402},
  numpages = {4},
  year = {2007},
  month = {Jul},
  publisher = {American Physical Society},
  doi = {10.1103/PhysRevLett.99.030402},
  urrl = {https://link.aps.org/doi/10.1103/PhysRevLett.99.030402}
}

@article{MCTDHB1,
  title = {Multiconfigurational time-dependent {Hartree} method for bosons: Many-body dynamics of bosonic systems},
  author = {Alon, Ofir E. and Streltsov, Alexej I. and Cederbaum, Lorenz S.},
  journal = {Phys. Rev. A},
  volume = {77},
  issue = {3},
  pages = {033613},
  numpages = {14},
  year = {2008},
  month = {Mar},
  publisher = {American Physical Society},
  doi = {10.1103/PhysRevA.77.033613},
  urrl = {https://link.aps.org/doi/10.1103/PhysRevA.77.033613}
}

@article{mctdhb_exact3,
  title = {Exact Quantum Dynamics of a Bosonic {Josephson} Junction},
  author = {Sakmann, Kaspar and Streltsov, Alexej I. and Alon, Ofir E. and Cederbaum, Lorenz S.},
  journal = {Phys. Rev. Lett.},
  volume = {103},
  issue = {22},
  pages = {220601},
  numpages = {4},
  year = {2009},
  month = {Nov},
  publisher = {American Physical Society},
  doi = {10.1103/PhysRevLett.103.220601},
  urrl = {https://link.aps.org/doi/10.1103/PhysRevLett.103.220601}
}

@Article{phantom_angular_axel,
author={Weiner, Storm E.
and Tsatsos, Marios C.
and Cederbaum, Lorenz S.
and Lode, Axel U. J.},
title={Phantom vortices: hidden angular momentum in ultracold dilute {Bose-Einstein} condensates},
journal={Sci. Rep.},
year={2017},
month={Jan},
day={16},
volume={7},
number={1},
pages={40122},
isssn={2045-2322},
doi={10.1038/srep40122},
urrl={https://doi.org/10.1038/srep40122}
}

@article{beinke_rotation1,
  title = {Many-body tunneling dynamics of {Bose-Einstein} condensates and vortex states in two spatial dimensions},
  author = {Beinke, Raphael and Klaiman, Shachar and Cederbaum, Lorenz S. and Streltsov, Alexej I. and Alon, Ofir E.},
  journal = {Phys. Rev. A},
  volume = {92},
  issue = {4},
  pages = {043627},
  numpages = {10},
  year = {2015},
  month = {Oct},
  publisher = {American Physical Society},
  doi = {10.1103/PhysRevA.92.043627},
  urrl = {https://link.aps.org/doi/10.1103/PhysRevA.92.043627}
}

@article{beinke_rotation2,
  title = {Enhanced many-body effects in the excitation spectrum of a weakly interacting rotating {Bose-Einstein} condensate},
  author = {Beinke, Raphael and Cederbaum, Lorenz S. and Alon, Ofir E.},
  journal = {Phys. Rev. A},
  volume = {98},
  issue = {5},
  pages = {053634},
  numpages = {9},
  year = {2018},
  month = {Nov},
  publisher = {American Physical Society},
  doi = {10.1103/PhysRevA.98.053634},
  urrl = {https://link.aps.org/doi/10.1103/PhysRevA.98.053634}
}

@article{mctdhb_review,
  title = {Colloquium: Multiconfigurational time-dependent {Hartree} approaches for indistinguishable particles},
  author = {Lode, Axel U. J. and L\'ev\^eque, Camille and Madsen, Lars Bojer and Streltsov, Alexej I. and Alon, Ofir E.},
  journal = {Rev. Mod. Phys.},
  volume = {92},
  issue = {1},
  pages = {011001},
  numpages = {21},
  year = {2020},
  month = {Feb},
  publisher = {American Physical Society},
  doi = {10.1103/RevModPhys.92.011001},
  urrl = {https://link.aps.org/doi/10.1103/RevModPhys.92.011001}
}

@article{paolo_fermion,
  title = {Stability of quasicrystalline ultracold fermions to dipolar interactions},
  author = {Molignini, Paolo},
  journal = {Phys. Rev. Res.},
  volume = {7},
  issue = {3},
  pages = {L032026},
  numpages = {9},
  year = {2025},
  month = {Aug},
  publisher = {American Physical Society},
  doi = {10.1103/szdc-61nl},
  url = {https://link.aps.org/doi/10.1103/szdc-61nl}
}

@article{paolo_cavity,
  title = {Superfluid--{Mott}-insulator transition of ultracold superradiant bosons in a cavity},
  author = {Lin, Rui and Papariello, Luca and Molignini, Paolo and Chitra, R. and Lode, Axel U. J.},
  journal = {Phys. Rev. A},
  volume = {100},
  issue = {1},
  pages = {013611},
  numpages = {15},
  year = {2019},
  month = {Jul},
  publisher = {American Physical Society},
  doi = {10.1103/PhysRevA.100.013611},
  urrl = {https://link.aps.org/doi/10.1103/PhysRevA.100.013611}
}

@article{paolo_ol,
  title = {Accuracy of quantum simulators with ultracold dipolar molecules: A quantitative comparison between continuum and lattice descriptions},
  author = {Hughes, Michael and Lode, Axel U. J. and Jaksch, Dieter and Molignini, Paolo},
  journal = {Phys. Rev. A},
  volume = {107},
  issue = {3},
  pages = {033323},
  numpages = {10},
  year = {2023},
  month = {Mar},
  publisher = {American Physical Society},
  doi = {10.1103/PhysRevA.107.033323},
  urrl = {https://link.aps.org/doi/10.1103/PhysRevA.107.033323}
}

@article{paolo_cz,
  title = {Crystallization via cavity-assisted infinite-range interactions},
  author = {Molignini, Paolo and L\'ev\^eque, Camille and Ke\ss{}ler, Hans and Jaksch, Dieter and Chitra, R. and Lode, Axel U. J.},
  journal = {Phys. Rev. A},
  volume = {106},
  issue = {1},
  pages = {L011701},
  numpages = {7},
  year = {2022},
  month = {Jul},
  publisher = {American Physical Society},
  doi = {10.1103/PhysRevA.106.L011701},
  urrl = {https://link.aps.org/doi/10.1103/PhysRevA.106.L011701}
}

@article{paolo_rotation,
doi = {10.1088/1361-648X/ae0fd3},
url = {https://doi.org/10.1088/1361-648X/ae0fd3},
year = {2025},
month = {nov},
publisher = {IOP Publishing},
volume = {37},
number = {44},
pages = {445401},
author = {Molignini, Paolo},
title = {Beyond-mean-field phases of rotating dipolar condensates in the strongly correlated regime},
journal = {J. Phys.: Cond. Matt.}
}

@Article{paolo_mott,
	title={{Mott transition in a cavity-boson system: A quantitative comparison between theory and experiment}},
	author={Rui Lin and Christoph Georges and Jens Klinder and Paolo Molignini and Miriam Büttner and Axel U. J. Lode and R. Chitra and Andreas Hemmerich and Hans Keßler},
	journal={SciPost Phys.},
	volume={11},
	pages={030},
	year={2021},
	publisher={SciPost},
	doi={10.21468/SciPostPhys.11.2.030},
	urrl={https://scipost.org/10.21468/SciPostPhys.11.2.030},
}

@article{paolo_cavity2,
  title = {Pathway to chaos through hierarchical superfluidity in blue-detuned cavity-{BEC} systems},
  author = {Lin, Rui and Molignini, Paolo and Lode, Axel U. J. and Chitra, R.},
  journal = {Phys. Rev. A},
  volume = {101},
  issue = {6},
  pages = {061602},
  numpages = {6},
  year = {2020},
  month = {Jun},
  publisher = {American Physical Society},
  doi = {10.1103/PhysRevA.101.061602},
  urrl = {https://link.aps.org/doi/10.1103/PhysRevA.101.061602}
}

@article{paolo_PhysRevA.98.053620,
  title = {Superlattice switching from parametric instabilities in a driven-dissipative {Bose-Einstein} condensate in a cavity},
  author = {Molignini, Paolo and Papariello, Luca and Lode, Axel U. J. and Chitra, R.},
  journal = {Phys. Rev. A},
  volume = {98},
  issue = {5},
  pages = {053620},
  numpages = {9},
  year = {2018},
  month = {Nov},
  publisher = {American Physical Society},
  doi = {10.1103/PhysRevA.98.053620},
  urrl = {https://link.aps.org/doi/10.1103/PhysRevA.98.053620}
}

@article{paolo_shake_potential,
	title={{Pauli crystal melting in shaken optical traps}},
	author={Jiabing Xiang and Paolo Molignini and Miriam Büttner and Axel U. J. Lode},
	journal={SciPost Phys.},
	volume={14},
	pages={003},
	year={2023},
	publisher={SciPost},
	doi={10.21468/SciPostPhys.14.1.003},
	url={https://scipost.org/10.21468/SciPostPhys.14.1.003},
}

@article{ref1,
author = {N.R. Cooper},
title = {Rapidly rotating atomic gases},
journal = {Adv. Phys.},
volume = {57},
number = {6},
pages = {539},
year = {2008},
publisher = {Taylor \& Francis},
doi = {10.1080/00018730802564122},
URL = {https://doi.org/10.1080/00018730802564122}
}

@article{ref2,
  title = {Vortex and fractional quantum Hall phases in a rotating anisotropic {Bose} gas},
  author = {Tanyeri, U. and Kallushi, A. and Umucal\ifmmode \imath \else \i \fi{}lar, R. O. and Kele\ifmmode \mbox{\c{s}}\else \c{s}\fi{}, A.},
  journal = {Phys. Rev. A},
  volume = {112},
  issue = {2},
  pages = {023321},
  numpages = {10},
  year = {2025},
  month = {Aug},
  publisher = {American Physical Society},
  doi = {10.1103/bsvm-vgnb},
  url = {https://link.aps.org/doi/10.1103/bsvm-vgnb}
}

@article{ref3,
  title = {Supersonic Rotation of a Superfluid: A Long-Lived Dynamical Ring},
  author = {Guo, Yanliang and Dubessy, Romain and de Herve, Mathieu de Go\"er and Kumar, Avinash and Badr, Thomas and Perrin, Aur\'elien and Longchambon, Laurent and Perrin, H\'el\`ene},
  journal = {Phys. Rev. Lett.},
  volume = {124},
  issue = {2},
  pages = {025301},
  numpages = {5},
  year = {2020},
  month = {Jan},
  publisher = {American Physical Society},
  doi = {10.1103/PhysRevLett.124.025301},
  url = {https://link.aps.org/doi/10.1103/PhysRevLett.124.025301}
}

@article{ref4,
  title = {Rotating quantum turbulence in the unitary Fermi gas},
  author = {Hossain, Khalid and Kobuszewski, Konrad and Forbes, Michael McNeil and Magierski, Piotr and Sekizawa, Kazuyuki and Wlaz\l{}owski, Gabriel},
  journal = {Phys. Rev. A},
  volume = {105},
  issue = {1},
  pages = {013304},
  numpages = {11},
  year = {2022},
  month = {Jan},
  publisher = {American Physical Society},
  doi = {10.1103/PhysRevA.105.013304},
  url = {https://link.aps.org/doi/10.1103/PhysRevA.105.013304}
}

@article{ref5,
  title = {Fractional quantum Hall physics and higher-order momentum correlations in a few spinful fermionic contact-interacting ultracold atoms in rotating traps},
  author = {Yannouleas, Constantine and Landman, Uzi},
  journal = {Phys. Rev. A},
  volume = {102},
  issue = {4},
  pages = {043317},
  numpages = {23},
  year = {2020},
  month = {Oct},
  publisher = {American Physical Society},
  doi = {10.1103/PhysRevA.102.043317},
  url = {https://link.aps.org/doi/10.1103/PhysRevA.102.043317}
}

@Article{ref6,
author={L{\'e}onard, Julian
and Kim, Sooshin
and Kwan, Joyce
and Segura, Perrin
and Grusdt, Fabian
and Repellin, C{\'e}cile
and Goldman, Nathan
and Greiner, Markus},
title={Realization of a fractional quantum Hall state with ultracold atoms},
journal={Nature},
year={2023},
month={Jul},
day={01},
volume={619},
number={7970},
pages={495},
issn={1476-4687},
doi={10.1038/s41586-023-06122-4},
url={https://doi.org/10.1038/s41586-023-06122-4}
}

@article{ref7,
  title = {Realization of a Laughlin State of Two Rapidly Rotating Fermions},
  author = {Lunt, Philipp and Hill, Paul and Reiter, Johannes and Preiss, Philipp M. and Ga\l{}ka, Maciej and Jochim, Selim},
  journal = {Phys. Rev. Lett.},
  volume = {133},
  issue = {25},
  pages = {253401},
  numpages = {7},
  year = {2024},
  month = {Dec},
  publisher = {American Physical Society},
  doi = {10.1103/PhysRevLett.133.253401},
  url = {https://link.aps.org/doi/10.1103/PhysRevLett.133.253401}
}

@article{ref8,
doi = {10.1088/0953-4075/39/10/S09},
url = {https://doi.org/10.1088/0953-4075/39/10/S09},
year = {2006},
month = {may},
publisher = {},
volume = {39},
number = {10},
pages = {S89},
author = {Regnault, N and Chang, C C and Jolicoeur, Th and Jain, J K},
title = {Composite fermion theory of rapidly rotating two-dimensional bosons},
journal = {J. Phys. B: At., Mol. and Opt. Phys.}
}

@article{app1,
author = {Zembaty, Zbigniew and Bernauer, Felix and Igel, Heiner and Schreiber, Karl Ulrich},
title = {Rotation Rate Sensors and Their Applications},
journal = {Sensors},
VOLUME = {21},
YEAR = {2021},
number = {16},
pages = {5344},
url = {https://www.mdpi.com/1424-8220/21/16/5344},
ISSN = {1424-8220},
doi = {10.3390/s21165344}
}

@book{app2,
author = {Farrell, Jay},
title = {Aided Navigation: GPS with High Rate Sensors},
year = {2008},
isbn = {0071493298},
publisher = {McGraw-Hill, Inc.},
address = {USA},
edition = {1}
}

@Article{app3,
AUTHOR = {Sollberger, David and Igel, Heiner and Schmelzbach, Cedric and Edme, Pascal and van Manen, Dirk-Jan and Bernauer, Felix and Yuan, Shihao and Wassermann, Joachim and Schreiber, Ulrich and Robertsson, Johan O. A.},
TITLE = {Seismological Processing of Six Degree-of-Freedom Ground-Motion Data},
JOURNAL = {Sensors},
VOLUME = {20},
YEAR = {2020},
NUMBER = {23},
PAGES = {6904},
URL = {https://www.mdpi.com/1424-8220/20/23/6904},
PubMedID = {33287180},
ISSN = {1424-8220},
DOI = {10.3390/s20236904}
}

@Article{app4,
AUTHOR = {Luo, Hang and Su, Hongbin and Tang, Qiwen and Nisa, Fazal ul and He, Liang and Zhang, Tao and Liu, Xiaoyu and Liu, Zhen},
TITLE = {Review of Research Advances in Gyroscopes’ Structural Forms and Processing Technologies Viewed from Performance Indices},
JOURNAL = {Sensors},
VOLUME = {25},
YEAR = {2025},
NUMBER = {19},
PAGES = {6193},
URL = {https://www.mdpi.com/1424-8220/25/19/6193},
PubMedID = {41095013},
ISSN = {1424-8220},
DOI = {10.3390/s25196193}
}

@article{app11,
  title = {The ring laser gyro},
  author = {Chow, W. W. and Gea-Banacloche, J. and Pedrotti, L. M. and Sanders, V. E. and Schleich, W. and Scully, M. O.},
  journal = {Rev. Mod. Phys.},
  volume = {57},
  issue = {1},
  pages = {61},
  numpages = {0},
  year = {1985},
  month = {Jan},
  publisher = {American Physical Society},
  doi = {10.1103/RevModPhys.57.61},
  url = {https://link.aps.org/doi/10.1103/RevModPhys.57.61}
}

@article{app12,
author = {H. J. Arditty and H. C. Lef\`{e}vre},
journal = {Opt. Lett.},
number = {8},
pages = {401},
publisher = {Optica Publishing Group},
title = {Sagnac effect in fiber gyroscopes},
volume = {6},
month = {Aug},
year = {1981},
url = {https://opg.optica.org/ol/abstract.cfm?URI=ol-6-8-401},
doi = {10.1364/OL.6.000401}
}

@article{app14,
  title = {Noise Level of a Ring Laser Gyroscope in the Femto-Rad/s Range},
  author = {Di Virgilio, Angela D. V. and Bajardi, Francesco and Basti, Andrea and Beverini, Nicol\`o and Carelli, Giorgio and Ciampini, Donatella and Di Somma, Giuseppe and Fuso, Francesco and Maccioni, Enrico and Marsili, Paolo and Ortolan, Antonello and Porzio, Alberto and Vitali, David},
  journal = {Phys. Rev. Lett.},
  volume = {133},
  issue = {1},
  pages = {013601},
  numpages = {6},
  year = {2024},
  month = {Jul},
  publisher = {American Physical Society},
  doi = {10.1103/PhysRevLett.133.013601},
  url = {https://link.aps.org/doi/10.1103/PhysRevLett.133.013601}
}

@article{app15,
  title = {Quantum Rotation Sensing with Dual Sagnac Interferometers in an Atom-Optical Waveguide},
  author = {Moan, E. R. and Horne, R. A. and Arpornthip, T. and Luo, Z. and Fallon, A. J. and Berl, S. J. and Sackett, C. A.},
  journal = {Phys. Rev. Lett.},
  volume = {124},
  issue = {12},
  pages = {120403},
  numpages = {5},
  year = {2020},
  month = {Mar},
  publisher = {American Physical Society},
  doi = {10.1103/PhysRevLett.124.120403},
  url = {https://link.aps.org/doi/10.1103/PhysRevLett.124.120403}
}

@Article{app16,
author={Schubert, Christian
and Abend, Sven
and Gersemann, Matthias
and Gebbe, Martina
and Schlippert, Dennis
and Berg, Peter
and Rasel, Ernst M.},
title={Multi-loop atomic Sagnac interferometry},
journal={Sci. Rep.},
year={2021},
month={Aug},
day={09},
volume={11},
number={1},
pages={16121},
issn={2045-2322},
doi={10.1038/s41598-021-95334-7},
url={https://doi.org/10.1038/s41598-021-95334-7}
}

@article{app21,
author = {Yuwen Cao and Xiangdong Ma and Yanjun Chen and Huimin Huang and Lanxin Zhu and Wenbo Wang and Yan He and Zhengbin Li},
journal = {Opt. Lett.},
number = {11},
pages = {3067},
publisher = {Optica Publishing Group},
title = {Interferometric fiber-optic gyroscope based on mode-division multiplexing},
volume = {48},
month = {Jun},
year = {2023},
url = {https://opg.optica.org/ol/abstract.cfm?URI=ol-48-11-3067},
doi = {10.1364/OL.494087}
}

@Article{app22,
AUTHOR = {Choi, Woo-Seok and Shim, Kyu-Min and Chong, Kyung-Ho and An, Jun-Eon and Kim, Cheon-Joong and Park, Byung-Yoon},
TITLE = {Sagnac Effect Compensations and Locked States in a Ring Laser Gyroscope},
JOURNAL = {Sensors},
VOLUME = {23},
YEAR = {2023},
NUMBER = {3},
PAGES = {1718},
URL = {https://www.mdpi.com/1424-8220/23/3/1718},
PubMedID = {36772752},
ISSN = {1424-8220},
DOI = {10.3390/s23031718}
}

@Article{app31,
AUTHOR = {Chang, Honglong and Xue, Liang and Qin, Wei and Yuan, Guangmin and Yuan, Weizheng},
TITLE = {An Integrated {MEMS} Gyroscope Array with Higher Accuracy Output},
JOURNAL = {Sensors},
VOLUME = {8},
YEAR = {2008},
NUMBER = {4},
PAGES = {2886},
URL = {https://www.mdpi.com/1424-8220/8/4/2886},
PubMedID = {27879855},
ISSN = {1424-8220},
DOI = {10.3390/s8042886}
}

@article{app32,
    author = {Guo, Xiaoting and Sun, Changku and Wang, Peng and Huang, Lu},
    title = {Vision sensor and dual {MEMS} gyroscope integrated system for attitude determination on moving base},
    journal = {Rev. Sci. Instrum.},
    volume = {89},
    number = {1},
    pages = {015002},
    year = {2018},
    month = {01},
    issn = {0034-6748},
    doi = {10.1063/1.5011703},
    url = {https://doi.org/10.1063/1.5011703}
}

@article{app33,
title = {Analytical and experimental study of stress effects in a {MEMS} ring gyroscope},
journal = {Sens. Actuators A Phys.},
volume = {362},
pages = {114639},
year = {2023},
issn = {0924-4247},
doi = {https://doi.org/10.1016/j.sna.2023.114639},
url = {https://www.sciencedirect.com/science/article/pii/S0924424723004880},
author = {Mehran Hosseini-Pishrobat and Derin Erkan and Erdinc Tatar}
}

@ARTICLE{app34,
title    = {Consideration of {Thermo-Vacuum} Stability of a {MEMS} Gyroscope for Space Applications},
author   = {Liu, Jili and Fu, Mingrui and Meng, Chao and Li, Jianpeng and Li, Kai and Hu, Jun and Chen, Xiaojuan},
journal  = {Sensors (Basel)},
volume   =  {20},
number   =  {24},
month    =  {Dec},
year     =  {2020}
}

@article{app41,
  title = {Coriolis-force-induced coupling between two modes of a mechanical resonator for detection of angular velocity},
  author = {Li, Kai and Fu, Hao and Li, Yong},
  journal = {Phys. Rev. A},
  volume = {98},
  issue = {2},
  pages = {023862},
  numpages = {6},
  year = {2018},
  month = {Aug},
  publisher = {American Physical Society},
  doi = {10.1103/PhysRevA.98.023862},
  url = {https://link.aps.org/doi/10.1103/PhysRevA.98.023862}
}

@article{app42,
  title = {Nuclear Spin Gyroscope based on the Nitrogen Vacancy Center in Diamond},
  author = {Soshenko, Vladimir V. and Bolshedvorskii, Stepan V. and Rubinas, Olga and Sorokin, Vadim N. and Smolyaninov, Andrey N. and Vorobyov, Vadim V. and Akimov, Alexey V.},
  journal = {Phys. Rev. Lett.},
  volume = {126},
  issue = {19},
  pages = {197702},
  numpages = {6},
  year = {2021},
  month = {May},
  publisher = {American Physical Society},
  doi = {10.1103/PhysRevLett.126.197702},
  url = {https://link.aps.org/doi/10.1103/PhysRevLett.126.197702}
}

@article{app43,
author = {Raffaele Silvestri  and Haocun Yu  and Teodor Strömberg  and Christopher Hilweg  and Robert W. Peterson  and Philip Walther },
title = {Experimental observation of Earth’s rotation with quantum entanglement},
journal = {Sci. Adv.},
volume = {10},
number = {24},
pages = {eado0215},
year = {2024},
doi = {10.1126/sciadv.ado0215},
URL = {https://www.science.org/doi/abs/10.1126/sciadv.ado0215}}

@article{app44,
author = {Clément Salducci  and Yannick Bidel  and Malo Cadoret  and Sarah Darmon  and Nassim Zahzam  and Alexis Bonnin  and Sylvain Schwartz  and Cédric Blanchard  and Alexandre Bresson },
title = {Quantum sensing of acceleration and rotation by interfering magnetically launched atoms},
journal = {Sci. Adv.},
volume = {10},
number = {44},
pages = {eadq4498},
year = {2024},
doi = {10.1126/sciadv.adq4498},
URL = {https://www.science.org/doi/abs/10.1126/sciadv.adq4498}}

@Article{app51,
author={Schumm, T.
and Hofferberth, S.
and Andersson, L. M.
and Wildermuth, S.
and Groth, S.
and Bar-Joseph, I.
and Schmiedmayer, J.
and Kr{\"u}ger, P.},
title={Matter-wave interferometry in a double well on an atom chip},
journal={Nat. Phys.},
year={2005},
month={Oct},
day={01},
volume={1},
number={1},
pages={57},
issn={1745-2481},
doi={10.1038/nphys125},
url={https://doi.org/10.1038/nphys125}
}

@Article{app52,
author={Hensel, T.
and Loriani, S.
and Schubert, C.
and Fitzek, F.
and Abend, S.
and Ahlers, H.
and Siem{\ss}, J.-N.
and Hammerer, K.
and Rasel, E. M.
and Gaaloul, N.},
title={Inertial sensing with quantum gases: a comparative performance study of condensed versus thermal sources for atom interferometry},
journal={Eur. Phys. J. D},
year={2021},
month={Mar},
day={22},
volume={75},
number={3},
pages={108},
issn={1434-6079},
doi={10.1140/epjd/s10053-021-00069-9},
url={https://doi.org/10.1140/epjd/s10053-021-00069-9}
}

@article{app53,
  title = {Optics and interferometry with atoms and molecules},
  author = {Cronin, Alexander D. and Schmiedmayer, J\"org and Pritchard, David E.},
  journal = {Rev. Mod. Phys.},
  volume = {81},
  issue = {3},
  pages = {1051},
  numpages = {0},
  year = {2009},
  month = {Jul},
  publisher = {American Physical Society},
  doi = {10.1103/RevModPhys.81.1051},
  url = {https://link.aps.org/doi/10.1103/RevModPhys.81.1051}
}

@article{app54,
doi = {10.1088/0034-4885/75/1/016401},
url = {https://doi.org/10.1088/0034-4885/75/1/016401},
year = {2011},
month = {dec},
publisher = {},
volume = {75},
number = {1},
pages = {016401},
author = {Sato, Y and Packard, R E},
title = {Superfluid helium quantum interference devices: physics and applications},
journal = {Rep. Prog. Phys.}
}

@article{app55,
  title = {Entanglement-Enhanced Atomic Gravimeter},
  author = {Cassens, Christophe and Meyer-Hoppe, Bernd and Rasel, Ernst and Klempt, Carsten},
  journal = {Phys. Rev. X},
  volume = {15},
  issue = {1},
  pages = {011029},
  numpages = {7},
  year = {2025},
  month = {Feb},
  publisher = {American Physical Society},
  doi = {10.1103/PhysRevX.15.011029},
  url = {https://link.aps.org/doi/10.1103/PhysRevX.15.011029}
}

@article{app56,
  title = {Acceleration-driven dynamics of {Josephson} vortices in coplanar superfluid rings},
  author = {Borysenko, Yurii and Bazhan, Nataliia and Prykhodko, Olena and Pfeiffer, Dominik and Lind, Ludwig and Birkl, Gerhard and Yakimenko, Alexander},
  journal = {Phys. Rev. A},
  volume = {111},
  issue = {4},
  pages = {043308},
  numpages = {10},
  year = {2025},
  month = {Apr},
  publisher = {American Physical Society},
  doi = {10.1103/PhysRevA.111.043308},
  url = {https://link.aps.org/doi/10.1103/PhysRevA.111.043308}
}

@article{app57,
  title = {High-Precision Quantum-Enhanced Gravimetry with a {Bose-Einstein} Condensate},
  author = {Szigeti, Stuart S. and Nolan, Samuel P. and Close, John D. and Haine, Simon A.},
  journal = {Phys. Rev. Lett.},
  volume = {125},
  issue = {10},
  pages = {100402},
  numpages = {8},
  year = {2020},
  month = {Sep},
  publisher = {American Physical Society},
  doi = {10.1103/PhysRevLett.125.100402},
  url = {https://link.aps.org/doi/10.1103/PhysRevLett.125.100402}
}

@article{rot_exp2,
author = {J. R. Abo-Shaeer  and C. Raman  and J. M. Vogels  and W. Ketterle },
title = {Observation of Vortex Lattices in {Bose-Einstein} Condensates},
journal = {Science},
volume = {292},
number = {5516},
pages = {476},
year = {2001},
doi = {10.1126/science.1060182},
URL = {https://www.science.org/doi/abs/10.1126/science.1060182}}

@article{rot_exp1,
  title = {Vortex Nucleation in a Stirred {Bose-Einstein} Condensate},
  author = {Raman, C. and Abo-Shaeer, J. R. and Vogels, J. M. and Xu, K. and Ketterle, W.},
  journal = {Phys. Rev. Lett.},
  volume = {87},
  issue = {21},
  pages = {210402},
  numpages = {4},
  year = {2001},
  month = {Nov},
  publisher = {American Physical Society},
  doi = {10.1103/PhysRevLett.87.210402},
  url = {https://link.aps.org/doi/10.1103/PhysRevLett.87.210402}
}

@article{MCTDHB_fermion1,
  title = {Correlated multielectron systems in strong laser fields: A multiconfiguration time-dependent Hartree-Fock approach},
  author = {Caillat, J. and Zanghellini, J. and Kitzler, M. and Koch, O. and Kreuzer, W. and Scrinzi, A.},
  journal = {Phys. Rev. A},
  volume = {71},
  issue = {1},
  pages = {012712},
  numpages = {13},
  year = {2005},
  month = {Jan},
  publisher = {American Physical Society},
  doi = {10.1103/PhysRevA.71.012712},
  url = {https://link.aps.org/doi/10.1103/PhysRevA.71.012712}
}

@article{MCTDHB_fermion2,
title = {Time-dependent multiconfiguration theory for electronic dynamics of molecules in an intense laser field},
journal = {Chem. Phys. Lett.},
volume = {392},
number = {4},
pages = {533},
year = {2004},
issn = {0009-2614},
doi = {https://doi.org/10.1016/j.cplett.2004.05.106},
url = {https://www.sciencedirect.com/science/article/pii/S0009261404008243},
author = {Tsuyoshi Kato and Hirohiko Kono}}

@article{MCTDHB_boson2,
  title = {Breaking the resilience of a two-dimensional {Bose-Einstein} condensate to fragmentation},
  author = {Klaiman, Shachar and Lode, Axel U. J. and Streltsov, Alexej I. and Cederbaum, Lorenz S. and Alon, Ofir E.},
  journal = {Phys. Rev. A},
  volume = {90},
  issue = {4},
  pages = {043620},
  numpages = {6},
  year = {2014},
  month = {Oct},
  publisher = {American Physical Society},
  doi = {10.1103/PhysRevA.90.043620},
  url = {https://link.aps.org/doi/10.1103/PhysRevA.90.043620}
}

@article{axel_spinor,
  title = {Multiconfigurational time-dependent {Hartree} method for bosons with internal degrees of freedom: Theory and composite fragmentation of multicomponent {Bose-Einstein} condensates},
  author = {Lode, Axel U. J.},
  journal = {Phys. Rev. A},
  volume = {93},
  issue = {6},
  pages = {063601},
  numpages = {10},
  year = {2016},
  month = {Jun},
  publisher = {American Physical Society},
  doi = {10.1103/PhysRevA.93.063601},
  url = {https://link.aps.org/doi/10.1103/PhysRevA.93.063601}
}

@article{Mistakidis_ol,
  title = {Many-body expansion dynamics of a {Bose-Fermi} mixture confined in an optical lattice},
  author = {Siegl, P. and Mistakidis, S. I. and Schmelcher, P.},
  journal = {Phys. Rev. A},
  volume = {97},
  issue = {5},
  pages = {053626},
  numpages = {12},
  year = {2018},
  month = {May},
  publisher = {American Physical Society},
  doi = {10.1103/PhysRevA.97.053626},
  url = {https://link.aps.org/doi/10.1103/PhysRevA.97.053626}
}

@article{Mistakidis_dw,
  title = {Many-body collisional dynamics of impurities injected into a double-well trapped {Bose-Einstein} condensate},
  author = {Theel, Friethjof and Keiler, Kevin and Mistakidis, Simeon I. and Schmelcher, Peter},
  journal = {Phys. Rev. Res.},
  volume = {3},
  issue = {2},
  pages = {023068},
  numpages = {16},
  year = {2021},
  month = {Apr},
  publisher = {American Physical Society},
  doi = {10.1103/PhysRevResearch.3.023068},
  url = {https://link.aps.org/doi/10.1103/PhysRevResearch.3.023068}
}

@article{Mistakidis_ol2,
title = {Bosonic quantum dynamics following a linear interaction quench in finite optical lattices of unit filling},
journal = {Chem. Phys.},
volume = {509},
pages = {106},
year = {2018},
issn = {0301-0104},
doi = {https://doi.org/10.1016/j.chemphys.2017.11.022},
url = {https://www.sciencedirect.com/science/article/pii/S0301010417306985},
author = {S.I. Mistakidis and G.M. Koutentakis and P. Schmelcher}}

@article{gammal_rotation,
  title = {Breakup of rotating asymmetric quartic-quadratic trapped condensates},
  author = {Brito, Leonardo and Andriati, Alex and Tomio, Lauro and Gammal, Arnaldo},
  journal = {Phys. Rev. A},
  volume = {102},
  issue = {6},
  pages = {063330},
  numpages = {13},
  year = {2020},
  month = {Dec},
  publisher = {American Physical Society},
  doi = {10.1103/PhysRevA.102.063330},
  url = {https://link.aps.org/doi/10.1103/PhysRevA.102.063330}
}

@article{axel_pelster1,
  title = {Thermodynamical properties of a rotating ideal {Bose} gas},
  author = {Kling, Sebastian and Pelster, Axel},
  journal = {Phys. Rev. A},
  volume = {76},
  issue = {2},
  pages = {023609},
  numpages = {6},
  year = {2007},
  month = {Aug},
  publisher = {American Physical Society},
  doi = {10.1103/PhysRevA.76.023609},
  url = {https://link.aps.org/doi/10.1103/PhysRevA.76.023609}
}

@Article{axel_pelster2,
author={Howe, K.
and Lima, A. R. P.
and Pelster, A.},
title={Rotating Fermi gases in an anharmonic trap},
journal={Eur. Phys. J. D},
year={2009},
month={Sep},
day={01},
volume={54},
number={3},
pages={667},
issn={1434-6079},
doi={10.1140/epjd/e2009-00182-9},
url={https://doi.org/10.1140/epjd/e2009-00182-9}
}

@article{kroha1,
  title = {Fluctuation damping of isolated, oscillating {Bose-Einstein} condensates},
  author = {Lappe, Tim and Posazhennikova, Anna and Kroha, Johann},
  journal = {Phys. Rev. A},
  volume = {98},
  issue = {2},
  pages = {023626},
  numpages = {12},
  year = {2018},
  month = {Aug},
  publisher = {American Physical Society},
  doi = {10.1103/PhysRevA.98.023626},
  url = {https://link.aps.org/doi/10.1103/PhysRevA.98.023626}
}

@article{kroha2,
  title = {Classical route to ergodicity and scarring phenomena in a two-component {Bose-Josephson} junction},
  author = {Mondal, Debabrata and Sinha, Sudip and Ray, Sayak and Kroha, Johann and Sinha, Subhasis},
  journal = {Phys. Rev. A},
  volume = {106},
  issue = {4},
  pages = {043321},
  numpages = {10},
  year = {2022},
  month = {Oct},
  publisher = {American Physical Society},
  doi = {10.1103/PhysRevA.106.043321},
  url = {https://link.aps.org/doi/10.1103/PhysRevA.106.043321}
}

@article{Thoss1,
    author = {Wang, Haobin and Pshenichnyuk, Ivan and Härtle, Rainer and Thoss, Michael},
    title = {Numerically exact, time-dependent treatment of vibrationally coupled electron transport in single-molecule junctions},
    journal = {J. Chem. Phys.},
    volume = {135},
    number = {24},
    pages = {244506},
    year = {2011},
    month = {12},
    issn = {0021-9606},
    doi = {10.1063/1.3660206},
    url = {https://doi.org/10.1063/1.3660206}
}

@article{Thoss2,
  title = {Nonequilibrium quantum systems with electron-phonon interactions: Transient dynamics and approach to steady state},
  author = {Wilner, Eli Y. and Wang, Haobin and Thoss, Michael and Rabani, Eran},
  journal = {Phys. Rev. B},
  volume = {89},
  issue = {20},
  pages = {205129},
  numpages = {14},
  year = {2014},
  month = {May},
  publisher = {American Physical Society},
  doi = {10.1103/PhysRevB.89.205129},
  url = {https://link.aps.org/doi/10.1103/PhysRevB.89.205129}
}

@article{Thoss3,
doi = {10.1088/1367-2630/10/11/115005},
url = {https://doi.org/10.1088/1367-2630/10/11/115005},
year = {2008},
month = {nov},
publisher = {},
volume = {10},
number = {11},
pages = {115005},
author = {Wang, Haobin and Thoss, Michael},
title = {From coherent motion to localization: dynamics of the spin-boson model at zero temperature},
journal = {New J. Phys.}
}

@article{collapse_revival,
  title = {Quantum dynamics of an atomic {Bose-Einstein} condensate in a double-well potential},
  author = {Milburn, G. J. and Corney, J. and Wright, E. M. and Walls, D. F.},
  journal = {Phys. Rev. A},
  volume = {55},
  issue = {6},
  pages = {4318--4324},
  numpages = {0},
  year = {1997},
  month = {Jun},
  publisher = {American Physical Society},
  doi = {10.1103/PhysRevA.55.4318},
  url = {https://link.aps.org/doi/10.1103/PhysRevA.55.4318}
}

@Article{main_ref1,
author={Lachmann, Maike D.
and Ahlers, Holger
and Becker, Dennis
and Dinkelaker, Aline N.
and Grosse, Jens
and Hellmig, Ortwin
and M{\"u}ntinga, Hauke
and Schkolnik, Vladimir
and Seidel, Stephan T.
and Wendrich, Thijs
and Wenzlawski, Andr{\'e}
and Carrick, Benjamin
and Gaaloul, Naceur
and L{\"u}dtke, Daniel
and Braxmaier, Claus
and Ertmer, Wolfgang
and Krutzik, Markus
and L{\"a}mmerzahl, Claus
and Peters, Achim
and Schleich, Wolfgang P.
and Sengstock, Klaus
and Wicht, Andreas
and Windpassinger, Patrick
and Rasel, Ernst M.},
title={Ultracold atom interferometry in space},
journal={Nat. Commun.},
year={2021},
month={Feb},
day={26},
volume={12},
number={1},
pages={1317},
issn={2041-1723},
doi={10.1038/s41467-021-21628-z},
url={https://doi.org/10.1038/s41467-021-21628-z}
}

@article{main_ref2,
    author = {Abend, Sven and Allard, Baptiste and Arnold, Aidan S. and Ban, Ticijana and Barry, Liam and Battelier, Baptiste and Bawamia, Ahmad and Beaufils, Quentin and Bernon, Simon and Bertoldi, Andrea and Bonnin, Alexis and Bouyer, Philippe and Bresson, Alexandre and Burrow, Oliver S. and Canuel, Benjamin and Desruelle, Bruno and Drougakis, Giannis and Forsberg, René and Gaaloul, Naceur and Gauguet, Alexandre and Gersemann, Matthias and Griffin, Paul F. and Heine, Hendrik and Henderson, Victoria A. and Herr, Waldemar and Kanthak, Simon and Krutzik, Markus and Lachmann, Maike D. and Lammegger, Roland and Magnes, Werner and Mileti, Gaetano and Mitchell, Morgan W. and Mottini, Sergio and Papazoglou, Dimitris and Pereira dos Santos, Franck and Peters, Achim and Rasel, Ernst and Riis, Erling and Schubert, Christian and Seidel, Stephan Tobias and Tino, Guglielmo M. and Van Den Bossche, Mathias and von Klitzing, Wolf and Wicht, Andreas and Witkowski, Marcin and Zahzam, Nassim and Zawada, Michał},
    title = {Technology roadmap for cold-atoms based quantum inertial sensor in space},
    journal = {AVS Quantum Sci.},
    volume = {5},
    number = {1},
    pages = {019201},
    year = {2023},
    month = {03},
    issn = {2639-0213},
    doi = {10.1116/5.0098119},
    url = {https://doi.org/10.1116/5.0098119},
    eeprint = {https://pubs.aip.org/avs/aqs/article-pdf/doi/10.1116/5.0098119/19823996/019201_1_online.pdf},
}

@article{main_ref3,
  title = {Matter-wave analog of a fiber-optic gyroscope},
  author = {Krzyzanowska, Katarzyna A. and Ferreras, Jorge and Ryu, Changhyun and Samson, Edward Carlo and Boshier, Malcolm G.},
  journal = {Phys. Rev. A},
  volume = {108},
  issue = {4},
  pages = {043305},
  numpages = {10},
  year = {2023},
  month = {Oct},
  publisher = {American Physical Society},
  doi = {10.1103/PhysRevA.108.043305},
  url = {https://link.aps.org/doi/10.1103/PhysRevA.108.043305}
}

@article{vittorio,
author = {Bellettini, Alice and Penna, Vittorio},
title = {Supercurrents and Tunneling in Massive Many-Vortex Necklaces and Star-Lattices},
journal = {Annalen der Physik},
volume = {537},
number = {12},
pages = {e00268},
doi = {https://doi.org/10.1002/andp.202500268},
url = {https://onlinelibrary.wiley.com/doi/abs/10.1002/andp.202500268},
year = {2025}
}

@misc{Ulbricht,
      title={Analytical description of collisional decoherence in a {BEC} double-well accelerometer}, 
      author={Kateryna Korshynska and Sebastian Ulbricht},
      year={2025},
      eprint={2508.01090},
      archivePrefix={arXiv},
      primaryClass={cond-mat.quant-gas},
      url={https://arxiv.org/abs/2508.01090}, 
}
\end{document}